\documentclass[journal, draftcls, one column, 12pt]{IEEEtran}
\usepackage{graphicx}
\newcommand{\RomanNumeralCaps}[1]
{\MakeUppercase{\romannumeral #1}}

\usepackage{amsmath,amssymb,amsfonts}

\ifCLASSINFOpdf
\else
\fi
	\newtheorem{assumption}{Assumption}
\newtheorem{theorem}{Theorem}	
\newtheorem{proposition}{Proposition}
\newtheorem{definition}{Definition}

\usepackage{paralist}

\begin{document}

\title{Dynamic Matching and Allocation of Tasks}
%
%
%

\author{Kartik~Ahuja 
	and~Mihaela~van der Schaar,~\IEEEmembership{Fellow,~IEEE}
	\thanks{K. Ahuja and Mihaela van der Schaar are with the Department
		of Electrical and Computer Engineering, University of California, Los Angeles,
		CA, 90095 USA e-mail: (see ahujak@ucla.edu, mihaela@ee.ucla.edu)}
}

\maketitle

\begin{abstract}
In many two-sided markets, the parties to be matched have incomplete information about their characteristics. We consider the settings where the parties engaged are extremely patient and are interested in long-term partnerships. Hence, once the final matches are determined, they persist for a long time. Each side has an opportunity to learn (some) relevant information about the other  before final matches are made. For instance, clients seeking workers to perform tasks often conduct interviews that require the workers to perform some tasks and thereby provide information to both sides.  The performance of a worker in such an interview- and hence the information revealed - depends both on the inherent characteristics of the worker and the task and also on the actions taken by the worker (e.g. the effort expended), which are not observed by the client. Thus there is moral hazard. 
Our goal is to derive a dynamic matching mechanism that facilitates learning on both sides before final matches are achieved  and ensures that the worker side does not have incentive to obscure learning of their characteristics through their actions. We derive such a mechanism that leads to final matching that achieve optimal performance (revenue) in equilibrium. We show that the equilibrium strategy  is long-run coalitionally stable, which means there is no subset of workers and clients that can gain by deviating from the equilibrium strategy. We derive all the results under the modeling assumption that the utilities of the agents are defined as limit of means of the utility obtained in each interaction. 
\end{abstract}


%
\IEEEpeerreviewmaketitle

\section{Introduction}

\textbf{Motivation.} 

The seminal work of Holmstr{\"o}m \cite{holmstrom1999managerial} analyzes how the career concerns of an individual, that is, the incentives to influence the current behavior of the individual and the ability of the future employers to learn about her and hence, the individual's future rewards, represent a significant force to explain the behaviors observed in many market environments. These career concerns also arise in many two-sided matching settings. For instance, in job recruitment markets, the workers desire to be matched with the  clients. In industries, the managers desire to be matched with tasks/divisions. In medical school internships, the medical students desire to get internships. In these setups, the workers have career concerns, as their performance plays a significant role in determining their matches/position in the future. Both sides are self-interested and do not have sufficient information about their own and the other side's characteristics. 
The interactions between the two sides are repeated in nature, and the learning influences the future opportunities. The learning  during each interaction also depends on the actions taken by the two sides (e.g. the effort exerted by the workers during the interview, or the effort exerted by the managers on the tasks), which are not directly observed. Thus there is moral hazard. There can be many possible ways to organize the interactions over time. For instance, in job recruitment, the management needs to decide how to organize the interviews;  in freelancing, the platform (such as Upwork) decides the matching rule. The matching mechanism should ensure that it \textit{facilitates learning} on both sides before final matches are achieved and that one side does not feel incentivized to \textit{obscure learning} on the other side through their actions. Despite the ubiquitous nature of settings with matching and learning, there is no systematic theory that models these environments and characterizes  the optimal mechanisms that lead to  desirable matching.

%
%
%
%

\textbf{Problem overview and contributions.}
In this work, we consider a  repeated matching setting with two sides: workers and clients. The clients and workers are extremely patient and both sides are interested in long-term partnerships. Hence, once the final matches are made these matches persist for a long period of time. All the clients and workers start with no knowledge about their characteristics (the productivities of the workers, the cost of exerting effort for the workers, and the revenue generated by the tasks).  Every time a worker is matched to a task for the client, she decides the amount of effort to exert; the effort is not observed by the client and thus there is moral hazard.  The client observes the output of the worker, which depends on both the productivity and the effort from the worker. Since the effort is not observed by the client, she cannot learn the worker's true productivity. The worker may feel incentivized to select actions to obscure the learning and achieve better matches in the future. The worker observes the payments made by the clients, the effort it exerted, and the cost for the effort exerted.  The observations by the worker help her learn about her own characteristics. Since the interactions are long-term in nature and the parties involved are extremely patient we use limit of the means utility functions for each individual. A few examples of certain settings that are similar to our model are  long-term hiring (e.g., job recruitment by companies, long-term contracts on freelancing platforms such as Upwork), job rotation \cite{ortega2001job}.



We outline the main objective of this work next.

a. Define coalitional stability for dynamic matching with learning under moral hazard.

b. Construct a dynamic matching mechanism to achieve the following objectives: i) ensure workers are not incentivized, where the incentives are measured in terms of long-run payoffs, to hinder learning through their actions, 
ii) maximize the long-run revenue while ensuring that individuals follow a coalitionally stable equilibrium strategy.

We propose a definition of coalitional stability for environments with dynamic matching with learning in the presence of moral hazard. We construct a simple mechanism that achieves  coalitional stability and in some cases also achieves the optimal revenue. The mechanism has an  initial assessment phase where each worker and client are matched exactly once \footnote{Since we consider limit of means utilities the cost incurred in assessment phase do not contribute to the long-run utility.} followed by a reporting phase where both sides report their preferences. In the final phase, the clients and workers are matched based on their preferences using the Gale-Shapley algorithm. There can be many alternate choices for the design of the mechanism. For instance, the mechanism might solely match the workers and clients based on the revenue generated/output generated and without use of reports \cite{xiao2014incentive} or the mechanism might ask the workers to report their characteristics instead of their preferences. These alternate choices suffer from different limitations (e.g., a mechanism that is based solely on outputs can incentivize workers to obscure learning) while our mechanism satisfies the desired properties.

\textbf{Prior work.} There are several ways to categorize works in the area of matching: matching with or without transfers, matching with complete or incomplete information (with or without learning), matching with self-interested or obedient participants, matching in the presence/absence of moral hazard and adverse selection. 
We do not describe the works in these categories separately. Instead,  in Table \RomanNumeralCaps{1}, we compare with a set  of representative works in each category.  Next, we broadly position our work with respect to the existing works and then describe the works that are closest to us.

In many real matching setups, the presence of incomplete information is natural. For instance, in labor markets and marriage markets the two sides to be matched do not know each other's characteristics. However, in these markets when the entities on the two sides are matched to interact (worker producing output for the clients in labor markets,  interaction during dating in marriage markets), they use the observations made in the interaction to learn about each other. The observations made often depend both on the characteristics and on the actions (effort in the worker-client setting) taken strategically during the interaction, which makes learning the characteristics separately non-trivial. The interaction of such a learning process (obscured by actions) and its impact on the matching has not been studied in the existing works.  


Our previous works, \cite{xiao2014incentive}, \cite{xiao2016}, have studied matching environments similar to our work where  both the costly unobservable effort (moral hazard) and  unknown types (adverse selection) play a major role. In \cite{xiao2014incentive}, the  workers are assumed to be bounded-rational as they optimize a proxy version of their utility as defined by the conjecture function, while in the present work the workers are rational, foresighted and maximize their long-run utilities.  In \cite{xiao2014incentive}, \cite{xiao2016}, there is no learning of the workers' and tasks' characteristics (along the equilibrium path). The model proposed in \cite{xiao2014incentive}, \cite{xiao2016}   only applies to environments where the productivity  of the worker does not vary across the tasks. In comparison, the model in this current work is more general and applies to general matching environments where the tasks can be heterogeneous and is thus more practical. In  \cite{xiao2014incentive}, \cite{xiao2016},  the equilibrium matching need not necessarily be efficient: no provable guarantees with regard to optimization of revenue are given. Moreover,  \cite{xiao2014incentive}, \cite{xiao2016}, do not provide any stability gurarantees, unlike our work.


\section{Dynamic matching mechanism design}

\label{sec2}

In this section, we first describe the model and problem formulation. 
We use $\boldsymbol{A}$ for a matrix, $\boldsymbol{A}(i,j)$ for an element of the matrix, $\boldsymbol{a}$ to represent a vector,
$\boldsymbol{a}(i)$ for the $i^{th}$ element of the vector,  $\mathcal{A}$ for a set, and $a/A$ for a scalar.

\subsection{Model and problem formulation}
There is one planner, $N$ clients and $N$ workers who desire to be matched.\footnote{The entire analysis can be extended to the setting when the number of clients and workers is not equal.}  We define the set of $N$ workers as $\mathcal{N}=\{1,...,N\}$ and the set of tasks as $\mathcal{S}=\{1,..,N\}$.  Both the clients and workers are interested in long-term partnerships. Hence, each client and worker wants to find a match and continue working together.  We consider a discrete time infinite horizon model. We write each discrete time slot as $t\in \{0,1,...,\infty\}$.  
Each client has one task that it wants to be repeatedly executed in each time slot. The clients and workers are assumed to be rational. In each time slot, the clients and workers are assessed and matched according to the matching rule explained later. We  assume that in each time slot one worker can be matched to at most one client and vice-versa (one-to-one matching).   

\textbf{Quality distribution of the tasks.}
Each task is characterized by its quality level, which is equal to the revenue generated per unit of the task. $g:\mathcal{S}\rightarrow [g^{min},g^{max}]$  maps each task to its quality level of the task, where $ g^{min}>0$. We assume that $g$ is a strictly increasing function without loss of generality. We assume that the quality of the tasks is not known to anyone. 


\textbf{Productivity distribution of the workers.}   Each worker $i$'s productivity  is a measure of her skill level; it is the number of units of task a worker can complete per unit time. The productivity depends on both the worker and the type of the task that she performs. $\boldsymbol{F} : \mathcal{N} \times \mathcal{S} \rightarrow [f^{min}, f^{max}]$ is a mapping from every combination of worker and task to a productivity  level. We assume that no two workers have the same productivity  for a particular task $x$, which means $\boldsymbol{F}(i,x)=\boldsymbol{F}(k,x) \implies i=k$. We assume that the productivity  of the worker in performing a task is not known to anyone. (In Upwork, 96\% of the workers have no significant  experience \cite{tran2014efficient} to know their productivities).


\textbf{Efforts and outputs of the workers.} Each worker $i$ decides (strategically) how much effort $e_i$ to exert (time invested in working) on a task $x$, which is assigned in a particular time slot.  We assume that $e_i\in \mathcal{E}_{ix}=\{0,\delta,2\delta,..e_{ix}^{max}\}$, where $e_{ix}^{max}\in [e_{l}^{max}, e_{u}^{max}], \;\forall i\in \mathcal{N}, \;\forall x \in \mathcal{S}$.
The output produced,  which is the total number of units of  task $x$ completed, is given as $\boldsymbol{F}(i,x)e_i$ (speed of executing the task times the time spent working on it).  The effort exerted by a worker is  known privately to the worker only. The revenue generated is given as $\left[\boldsymbol{F}(i,x)e_{i}\right]g(x)$. We assume that the output produced and the revenue generated is observed by the client and the planner; this is a natural assumption, see \cite{holmstrom1999managerial}.  


We define a cost function $\boldsymbol{C}:\mathcal{S}\times \mathcal{N} \rightarrow [0, \infty) $.
It costs worker $i$  $\;\boldsymbol{C}(i,x)e_{i}^2$ to exert effort $e_{i}$ on task $x$, where $\boldsymbol{C}(i,x)\in [c^{min},c^{max}], \; \forall i \in \mathcal{N}, \forall x \in \mathcal{S}$. We assume a quadratic function here for simplifying the presentation; all the results extend to any convex cost function that increases in effort.  The worker $i$ does not know their own costs $\boldsymbol{C}(i,x),\;\forall x \in \mathcal{S}$ and no one else knows it as well. If worker $i$ is matched to a task $x$, then the worker observes the cost $\boldsymbol{C}(i,x)e_{i}^2$ and thus learns   $\boldsymbol{C}(i,x)$. Also, we define a constant $W^{max}=f^{max} \left[\max_{i\in \mathcal{N}, x\in \mathcal{S}} \{e_{ix}^{max}\}\right]$, which denotes the maximum output across all the workers. 

\textbf{Payment rule.}  We assume that the payment rules are fixed and the clients are required to follow the payment rules; only the concerned clients know the payment rules. In the Numerical Experiments Section, we discuss the scenario when the platform can optimize and prescribe from a set of payment rules.
In the extensions section and the Appendix, we discuss the client selected payment rules. 
If worker $i$ works on task $x$ and produces $\boldsymbol{W}(i,x)$ units of output (units of task completed by the worker), then the worker is paid  $p^F(\boldsymbol{W}(i,x), x) = \alpha \boldsymbol{W}(i,x)^2 g(x)$  by client $x$, where $\alpha$ is a given positive constant. \footnote{We choose a quadratic function for payments because the cost for exerting effort is quadratic.} We assume $\alpha$  to be less than $\frac{1}{2W^{max}}$ to guarantee a non-negative profit to all the clients (See the Appendix for details.). The payment rule is quadratic in the output of the worker to ensure proportional compensation of the quadratic costs for exerting effort. We can generalize the analysis to any form of payment rule (for instance, linear etc.); we provide details in the Appendix.

\textbf{Dynamic matching mechanisms.} 
The planner selects the matching rule and makes it public knowledge.  We first define a general vector of observations made by the planner up to time $t-1$ (end of time slot $t-1$) as $\boldsymbol{h}_{0}^{t}$. The elements of this general observation vector consist of the output histories of the workers, the actions that are taken by the workers (for instance, sending report about preferred clients to the planner), etc.


We define the set of all the possible histories of all possible lengths as $\mathcal{H}_{0} $.  A general matching rule is given as $\boldsymbol{m}: \mathcal{H}_{0} \rightarrow \Pi(\mathcal{S})$, where $\Pi(\mathcal{S})$ is the set of all possible permutations of $\mathcal{S}$. The matching rule maps each history of observations $\boldsymbol{h}_{0}^{t}$ to a vector of tasks. $\boldsymbol{m}(\boldsymbol{h}_0^{t})[i]$ denotes the $i^{th} $ element of the vector $\boldsymbol{m}(\boldsymbol{h}_0^{t})$ and corresponds to the task assigned to worker $i$ following history $\boldsymbol{h}_0^t$. 

In this work, as highlighted earlier, we are interested in settings where each individual wants to find a long-term match. Such situations arise in long-term contracts on platforms such as Upwork, job rotation \cite{ortega2001job}.  Therefore, we restrict ourselves to matching rules that satisfy the following condition:  $	\lim_{t\rightarrow \infty} \boldsymbol{m}(\boldsymbol{h}_0^{t})$ exists for all $\boldsymbol{h}_0^{t}\in \mathcal{H}_0$. Since these rules lead to a long-term match, we refer to these matching rules as \textit{long-term matching rules}. We denote the set of all long-term matching rules  as $\mathcal{M}$.  What about the matching rules for which the limits do not exist? This is true in the settings where the workers and clients do not engage in long-term contracts and instead work on short-term basis. For instance, on platforms such as Upwork the clients in some cases offer short-term contracts and not the long-term contracts.  We call the matching rules for which the limit do not exist as \textit{short-term matching rules}. Note that our analysis does not apply to these short-term matching rules and only applies to long-term matching rules.

\textbf{Strategies of the workers and clients.} We define a strategy as a mapping from the history of observations  to the actions.  We denote the strategies of the workers as $\{\boldsymbol{\pi}_{i}\}_{i=1}^{N}$ and the strategy for the clients as $\{\boldsymbol{\pi}_{i}\}_{i=N+1}^{2N}$. Each worker and client first need to decide whether or not to participate in the mechanism $\boldsymbol{m}$.  Each client and worker starts with no observation history (thus history at the start is $\phi$).  $\boldsymbol{\pi}_{i}(\phi) \in \{P, NP\}$ where $P$ is for participation and $NP$ is for not participation. Participation is the only active choice of a client  (In the extensions section, we discuss the client selected payment rules). If a set of clients or workers choose not to participate in the matching mechanism, they can pursue options outside the platform. But if they do participate, then they need to follow the matching mechanism set forth by the planner.

In each period, each worker decides to exert some effort on the task assigned, where the effort level is only known to the worker. In some mechanisms, the planner can solicit reports from the workers about their preference over different tasks. Each worker also observes the payments made and the costs incurred for exerting effort on the tasks.  We define the history of observations for each worker separately.  The vector of observations of a worker $i$ up to time $t$ as $\boldsymbol{h}_{i}^{t}$, which consists of the efforts exerted, reports sent, the payments received and the tasks assigned up to time slot $t-1$ (end of time slot $t-1$). In addition, $\boldsymbol{h}_{i}^{t}$  includes the task assigned in time slot $t$.    The  set of all the possible observations histories of all possible lengths is given as $\mathcal{H}_{i}$.
We define the strategy of worker $i$ as a mapping from the history of observations of the worker to the actions, $\boldsymbol{\pi}_{i}: \mathcal{H}_{i} \rightarrow \mathcal{A}_i$, where $\mathcal{A}_i$  is the set of actions that a worker takes.  $a_i\in\mathcal{A}_i$ has two components $a_i[1]$ is the effort exerted and $a_i[2]$ is the report vector. Different choices of $\boldsymbol{m}$  impact the action set differently. We define the set of all the possible strategies as $\Pi(\boldsymbol{m})$.



\textbf{The stage game.} 
In time slot $t$, worker $i$ is matched to play a stage game with client  $x=\boldsymbol{m}(\boldsymbol{h}_{0}^{t})[i]$ (assuming both agreed to participate in the mechanism).  The worker $i$  exerts $e_i^t$  effort following a private history $\boldsymbol{h}_{i}^{t}$ ($\boldsymbol{\pi}_{i}(\boldsymbol{h}_{i}^{t})[1]=e_i^{t}$). We define the output and the revenue generated  by worker $i$ in time slot $t$ for  client $x$ as
$W_{i}( \boldsymbol{h}_{0}^{t}, \boldsymbol{h}_{i}^{t}, \boldsymbol{\pi}_{i}|\boldsymbol{m}) = \boldsymbol{F}(i,x)e_i^{t}$ and $r_{i}( \boldsymbol{h}_{0}^{t}, \boldsymbol{h}_{i}^{t}, \boldsymbol{\pi}_{i}|\boldsymbol{m})=\boldsymbol{F}(i,x)g\left(x\right)e_i^{t}$ respectively. The payment made by  client $x$ to  worker $i$ for the corresponding output is given as $p(W_{i}( \boldsymbol{h}_{0}^{t}, \boldsymbol{h}_{i}^{t}, \boldsymbol{\pi}_{i}|\boldsymbol{m}), x)$. Therefore, the utility derived by the worker $i$ in the stage game played in time slot $t$ is computed as follows.
$
u_{i}(\boldsymbol{h}_{0}^{t}, \boldsymbol{h}_{i}^{t}, \boldsymbol{\pi}_i|\boldsymbol{m})=  p(W_{i}( \boldsymbol{h}_{0}^{t}, \boldsymbol{h}_{i}^{t}, \boldsymbol{\pi}_{i}|\boldsymbol{m}), x)  -\boldsymbol{C}(i,x)(e_i^t)^2
$

Note that the above utility is quasi-linear (linear in the payments). Similarly, the utility of  client $x$ (linear in the revenue and the payments made) who is matched to worker $i$ in time slot $t$ is given as follows.
$
v_{x}(\boldsymbol{h}_{0}^{t}, \boldsymbol{h}_{i}^{t}, \boldsymbol{\pi}_{i}|\boldsymbol{m}) = r_{i}( \boldsymbol{h}_{0}^{t}, \boldsymbol{h}_{i}^{t}, \boldsymbol{\pi}_{i}|\boldsymbol{m}) - p(W_{i}( \boldsymbol{h}_{0}^{t}, \boldsymbol{h}_{i}^{t}, \boldsymbol{\pi}_{i}|\boldsymbol{m}), x)
$

\textbf{The repeated endogenous matching game.} \label{rep-end-game} In every time slot, a stage game is played between a worker and a client who are matched \textit{endogenously} based on the observation history of the planner based on $\boldsymbol{m}$. We refer to this repeated game as the ``repeated endogenous matching game'' and define the long-run utility for each client and each worker next.

We assume that the workers and clients are extremely patient and define the long-run utility  for  worker $i$  and client $x$  as

$U_{i}(\{\boldsymbol{\pi}_{k}\}_{k=1}^{2N}|\boldsymbol{m})=\lim_{T\rightarrow \infty}\frac{1}{T+1}\sum_{t=0}^{T} u_{i}(\boldsymbol{h}_{0}^{t}, \boldsymbol{h}_{i}^{t}, \boldsymbol{\pi}_{i}|\boldsymbol{m}) $, 

$
V_{x}(\{\boldsymbol{\pi}_{k}\}_{k=1}^{2N}|\boldsymbol{m})= 	\lim_{T\rightarrow \infty} \frac{1}{T+1}\sum_{t=0}^{T}v_{x}(\boldsymbol{h}_{0}^{t}, \boldsymbol{h}_{i}^{t}, \boldsymbol{\pi}_{i}|\boldsymbol{m}) $ respectively. 

The total long-run revenue is  $R(\{\boldsymbol{\pi}_{k}\}_{k=1}^{2N}|\boldsymbol{m}) = \lim_{T\rightarrow \infty }\frac{1}{T+1}\sum_{t=0}^{T}\sum_{i=1}^{N}r_{i}( \boldsymbol{h}_{0}^{t}, \boldsymbol{h}_{i}^{t}, \boldsymbol{\pi}_{i}|\boldsymbol{m})$. It is fairly common to assume limit of means utilities in environments with long term career-concerns \cite{holmstrom1999managerial}. Also, the entire analysis  extends to discounted utilities (assuming the discount  is sufficiently high). 




%

\textbf{Knowledge and observation structure.}  The workers and the clients are rational, independent decision makers who do not cooperate in decision making and who wish to maximize their long-run utilities.  The payment rules  are known to the concerned client and the planner; the quality of the task is not known to anyone. The productivities  and the costs of exerting effort on a task for the workers are not known to anyone. The effort exerted by the worker and the corresponding set of effort levels are known to the worker privately. The structure of the utility (but not the parameters in the utility) of the workers and clients is known to the planner.  The output and the revenue produced by the worker is observed by the concerned client and the planner. The payment made by the client to the worker are observed by the worker, the client and the planner. The reports sent by the workers to the planner are kept private between the workers and the planner. This  knowledge  structure is common knowledge. We summarize the knowledge structure in Table \RomanNumeralCaps{2}.
\subsubsection{Long-run Stability of Matching}


We propose a definition of stability that extends the standard definitions to environments where dynamic matching is carried out with learning in the presence of moral hazard. 

%



Consider a matching rule $\boldsymbol{m}\in \mathcal{M}$. Suppose the joint strategy of all the workers and clients is given as $\boldsymbol{\pi}=\{\boldsymbol{\pi}_{1},...,\boldsymbol{\pi}_{2N}\}$. The history for the  planner induced by the joint strategy $\boldsymbol{\pi}$ is denoted as $\boldsymbol{h}_{0}^{t,\boldsymbol{\pi}}$ and the history for the worker $i$ induced by the joint strategy $\boldsymbol{\pi}$ is given as $\boldsymbol{h}_{i}^{t,\boldsymbol{\pi}}$.  The matching rule takes a limiting value depending upon the history, which we define as $\boldsymbol{m}^{*}_{\boldsymbol{\pi}} = \lim_{t\rightarrow \infty} \boldsymbol{m}(\boldsymbol{h}_{0}^{t,\boldsymbol{\pi}})$.    The expression for the long-run utilities for  worker $i$ and client $x=\boldsymbol{m}^{*}_{\boldsymbol{\pi}}[i] $ are simplified below (See details in the Appendix).

\begin{equation} 
U_{i}(\{\boldsymbol{\pi}_{k}\}_{k=1}^{2N}| \boldsymbol{m}) = \lim_{T\rightarrow \infty} \frac{1}{T+1} \sum_{t=0}^{T}\Big[p\big(\boldsymbol{F}(i, \boldsymbol{m}_{\boldsymbol{\pi}}^{*}[i])\boldsymbol{\pi}_{i}(\boldsymbol{h}_{i}^{t,\boldsymbol{\pi}}), \boldsymbol{m}_{\boldsymbol{\pi}}^{*}[i]\big)- \boldsymbol{C}(i, \boldsymbol{m}_{\boldsymbol{\pi}}^{*}[i])\boldsymbol{\pi}_{i}(\boldsymbol{h}_{i}^{t,\boldsymbol{\pi}})^2\Big]
\label{eqn-Uipi-simp}
\end{equation}
\begin{equation}V_{x}(\{\boldsymbol{\pi}_{k}\}_{k=1}^{2N}| \boldsymbol{m}) =\lim_{T\rightarrow \infty} \frac{1}{T+1} \sum_{t=0}^{T} \Big[\boldsymbol{F}(i, \boldsymbol{m}_{\boldsymbol{\pi}}^{*}[i])\boldsymbol{\pi}_{i}(\boldsymbol{h}_{i}^{t, \boldsymbol{\pi}}))g(j) - p\big(\boldsymbol{F}(i, \boldsymbol{m}_{\boldsymbol{\pi}}^{*}[i])\boldsymbol{\pi}_{i}(\boldsymbol{h}_{i}^{t,\boldsymbol{\pi}}), \boldsymbol{m}_{\boldsymbol{\pi}}^{*}[i]\big)\Big]
\label{eqn-Vxpi-simp}
\end{equation}

The above expression for long-run utility shows that the worker's utility depends on the task assigned in the limit and not on the utility derived in the phases before being matched to this task finally (thus any assessments/experimentation is costless). We now formalize the condition that no worker-task pair  that is not matched in $\boldsymbol{m}^{*}_{\boldsymbol{\pi}}$ cannot strictly gain by being matched to one another by jointly choosing to deviate from participating in the matching mechanism.
We assume that there are no side-payments, in other words, the payments are made solely based on worker's output following the given payment rule $p$.  Consider worker $i$ and a task $y$, where $y \not= \boldsymbol{m}^{*}_{\boldsymbol{\pi}}[i]$, and suppose that this worker-task pair is matched instead of $i$ and $\boldsymbol{m}_{\boldsymbol{\pi}}^{*}[i]$. In such a case, the long-run utilities achieved by worker $i$ and client $y$, when the strategy for worker $i$ is $\boldsymbol{\pi}^{'}_{i}$ and strategy for client $y$ is $\boldsymbol{\pi}^{'}_{y}$, is defined below in \eqref{eqn-Uipi-swap} and \eqref{eqn-Vypi-swap} respectively. Observe that we are considering a pairwise coalition of $i$ and task $y$ only. Therefore, the strategy of others cannot impact  worker $i$ and client $y$'s long-run utilities. Hence, it is sufficient to consider the strategy $\boldsymbol{\pi}_{i}^{'}$ to be a function of time only.


\begin{equation} \hat{U}_{i}(\boldsymbol{\pi}^{'}_{i}, \boldsymbol{\pi}^{'}_{y}) =  \lim_{T\rightarrow \infty} \frac{1}{T+1}\Big[ \sum_{t=0}^{T}p\Big(\boldsymbol{F}(i,y)\boldsymbol{\pi}_{i}^{'}(t),y\Big)- \boldsymbol{C}(i,y)\boldsymbol{\pi}_{i}^{'}(t)^2 \Big]
\label{eqn-Uipi-swap}
\end{equation}
\begin{equation}
\hat{V}_y(\boldsymbol{\pi}^{'}_{i},\boldsymbol{\pi}^{'}_{y}) = \lim_{T\rightarrow \infty} \frac{1}{T+1} \sum_{t=0}^{T}\boldsymbol{F}(i,y)\boldsymbol{\pi}_{i}^{'}(t)g(y) - p\Big(\boldsymbol{F}(i,y)\boldsymbol{\pi}_{i}^{'}(t), y\Big) 
\label{eqn-Vypi-swap}
\end{equation}

If a mechanism $\boldsymbol{m}$ is implemented, then we define long-run stability in terms of the above expressions for long-run pairwise utilities, \eqref{eqn-Uipi-simp}, \eqref{eqn-Vxpi-simp}, \eqref{eqn-Uipi-swap}, \eqref{eqn-Vypi-swap}  as follows. 

\begin{definition} \textit{Long-run Pairwise-Stability:} 
	A joint strategy $\boldsymbol{\pi}$ is long-run stable under $\boldsymbol{m}$ if there exists no worker-client pair $(i,y)$, not matched in the limit of $\boldsymbol{m}$ ($y\not=\boldsymbol{m}_{\boldsymbol{\pi}}^{*}(i) $), and a strategy for the  worker $i$ and  the client $y$  that leads to a strict increase in the long-run utility for both the worker $i$ and the client $y$, i.e. $ \hat{U}_{i}(\boldsymbol{\pi}^{'}_{i},\boldsymbol{\pi}^{'}_{y})> 
	U_{i}(\{\boldsymbol{\pi}_{k}\}_{k=1}^{2N}| \boldsymbol{m})$, $ \hat{V}_y(\boldsymbol{\pi}^{'}_{i}, \boldsymbol{\pi}^{'}_{y})> V_{y}(\{\boldsymbol{\pi}_{k}\}_{k=1}^{2N}|\boldsymbol{m})$.
\end{definition}

We extend the above definition from a pair of worker and client to any coalition of workers and clients. Suppose $\mathcal{X}$ is the set of the workers and clients who want to deviate and we define the strategy that an individual $i$ in the deviating set follows as $\boldsymbol{\pi}^{'}_{i}$.    The long-run utility of the worker $i \in \mathcal{X}$ (client $y \in \mathcal{X}$) is given as $\hat{U}_{i}(\{\boldsymbol{\pi}^{'}_{j}\}_{j \in \mathcal{X}}) $ ($\hat{V}_{y}(\{\boldsymbol{\pi}^{'}_{j}\}_{j \in \mathcal{X}}) $). 
\begin{definition} \textit{Long-run Coalitional-Stability:} 
	A joint strategy $\boldsymbol{\pi}$ is long-run coalition-stable under $\boldsymbol{m}$ if there exists no subset of workers and clients $\mathcal{X} \in \mathcal{S} \cup \mathcal{N}$,  and  strategies for the workers and clients in this subset given as $\{\boldsymbol{\pi}^{'}_{j}\}_{j \in \mathcal{X}}$ that leads to a strict increase in the long-run utility for each worker $i \in \mathcal{X}$ and client $y \in \mathcal{X}$, i.e. $ \hat{U}_{i}(\{\boldsymbol{\pi}^{'}_{j}\}_{j \in \mathcal{X}}) >
	U_{i}(\{\boldsymbol{\pi}_{k}\}_{k=1}^{2N}| \boldsymbol{m})$, $ \hat{V}_{y}(\{\pi^{'}_{j}\}_{j \in \mathcal{X}})> V_{y}(\{\boldsymbol{\pi}_{k}\}_{k=1}^{2N}|\boldsymbol{m})$.
\end{definition}

From the above it is clear that long-run coalitional-stability implies long-run pairwise-stability.

We now compare and contrast the difference of the proposed definition of long-run stability with the existing definitions.  Shapley's works \cite{gale1962college} and \cite{shapley1971assignment} proposed pairwise stability and core respectively.  More recently, there have been works on stability  in dynamic matching markets. In \cite{kadam2018multiperiod}, \cite{kennes2014day}, the authors analyze pairwise stability in  dynamic matching markets. In \cite{damiano2005stability}, \cite{kurino2009credibility}, \cite{doval2014theory}, authors analyze coalitional stability in dynamic matching markets. In our setup, unlike the existing setups, the preferences are learned  as there is incomplete information and the preferences are affected by the actions of one side.

%
%
%


\textbf{Planner's Problem.} 
The planner decides the mechanism $\boldsymbol{m}$ to maximize the total long-run revenue subject to three types of constraints. The first type of constraints are the individual rationality (IR) constraints, which if satisfied guarantee that the workers and the clients participate in the mechanism. The second type of constraints are the incentive-compatibility (IC) constraints, which guarantee that every worker follows an optimal strategy (given the strategies of others). If the strategy of each worker can satisfy the IC constraint, then the joint strategy of all the workers is an equilibrium (no worker will want to deviate).  We  also require that the joint strategy $\boldsymbol{\pi}$ is  long-run coalitionally stable under  $\boldsymbol{m}$. The planner's problem is 


\begin{equation*} 
\begin{split}
\max_{\boldsymbol{m} \in \mathcal{M}} & R(\{\boldsymbol{\pi}_{k}\}_{k=1}^{2N}|\boldsymbol{m}) \\
\text{s.t.}\;   & 	V_{x}(\{\boldsymbol{\pi}_{k}\}_{k=1}^{2N}|\boldsymbol{m}) \geq 0,\;  \forall x \in \mathcal{S}\; \text{(IR-clients)} \\
& U_{i}(\{\boldsymbol{\pi}_{k}\}_{k=1}^{2N}| \boldsymbol{m}) \geq 0\; \forall i \in \mathcal{N}\; \text{(IR-workers)}\\
& U_{i}(\boldsymbol{\pi}_i,\{\boldsymbol{\pi}_{k}\}_{k=1, k \not= i}^{2N}|\boldsymbol{m}) \geq U_{i}(\boldsymbol{\pi}_i^{'},\{\boldsymbol{\pi}_{k}\}_{k=1, k \not= i}^{2N}|\boldsymbol{m})\; \forall i \in \mathcal{N}\; \forall \boldsymbol{\pi}_{i}^{'}; \text{(IC-workers)} \\ 
& \boldsymbol{\pi} \; \text{is long-run coalition-stable under} \; \boldsymbol{m}\; 
\end{split}
\end{equation*}

The planner's problem outlined above \footnote{Long-run coalition-stability already implies that the other constraints in the planner's problem are satisfied but we still mention them separately for clarity. } is challenging because
\begin{itemize}
	
	\item \textbf{Incomplete information-} The planner needs to select $\boldsymbol{m}$ to maximize the total long-run revenue achieved by an equilibrium strategy, which depends on both the productivity  of the workers $\boldsymbol{F}$ and the costs $\boldsymbol{C}$ that are not known to the planner.   In our model, the planner and the workers do not even know the distribution of the workers' characteristics as is typically assumed in games of incomplete information to resolve the above challenge.
	
	
	\item \textbf{Computational intractability}- The sets of possible matching rules $\mathcal{M}$,  and the strategies of the workers $\Pi(\boldsymbol{m})$ is extremely large thus making the problem computationally intractable.
	

	%
\end{itemize}

%
%

%

\subsection{Proposed mechanism and its properties}
First, we give a brief description of the proposed mechanism. 
The proposed matching rule is designed to evaluate each worker on every type of task exactly once.  Since the worker is evaluated only once we refer to the proposed matching mechanism as ``first impression is the last impression'' (FILI). Based on the output of the workers a ranking of the workers over the different tasks is computed and the workers also submit a preference for the tasks to the planner. The planner computes a final matching based on these rankings and preferences, which remains fixed for all the future time slots. 


\textbf{Matching rule.}
The FILI matching rule denoted as $\boldsymbol{m}^F$ operates in three phases described below.
\begin{compactenum}
	
	\item \textbf{Assessment phase} ($0 \leq t\leq N-1$) Each worker is matched to every task exactly once in the first $N$ time slots. At the end of each time slot,  the worker, the client and the planner observe the output of the worker on the assigned task.  We write the observation of the planner in the form of a matrix $\boldsymbol{W}^{e}$, where $\boldsymbol{W}^{e}(i,x)$ is the output of worker $i$ on task $x$ in the assessment phase.
	

	\item \textbf{Reporting phase} ($t=N$) The planner requests all the workers to submit their preferences  in the form of ranks (strictly ordered) for tasks. These rank submissions are a part of the strategy for the workers, which we describe later.\footnote{In practical settings, not all the tasks on the platform are very different and many of them can be categorized into one type, for instance, translation (each worker has the same productivity  for tasks of the same type). In such cases, it is sufficient to evaluate the workers on tasks of different types.} The planner computes the preferences for the clients over the workers  based on the  outputs $\boldsymbol{W}^e$ as follows. For every client $x$, the planner ranks the workers based on the outputs produced on task $x\;$ $\{\boldsymbol{W}^e(i,x)\}_{i=1}^{N}$. The mechanism persists with the matching from the previous time slot for this time slot.

	\item \textbf{Operational phase ($t\geq N+1$)} 
	The planner computes the matching based on the G-S algorithm \cite{gale1962college} as follows. 
	The planner executes the G-S algorithm with the workers as the proposers and the clients as the acceptors. In each iteration of the algorithm, each worker proposes to her favorite task among the tasks that have not already rejected it. Each client based on the proposals it gets keeps its favorite worker on hold and rejects the rest.  At the end of at most $N^2-2N+2$  iterations,  the matching that is achieved is final. 
	The matching computed above is fixed for the remaining time slots starting from $N+1$. 
\end{compactenum}



Next, we state a proposition which shows that the workers and clients are always willing to participate in the above mechanism. 

\begin{proposition} \label{prop1}    It is individually rational for all the clients and the workers  to participate in the proposed mechanism.
\end{proposition}

The proofs of all the theorems and propositions are given in the Appendix. The proposed mechanism induces  a repeated endogenous matching game as described in Section \ref{rep-end-game}. In the next section, we derive an equilibrium strategy for this repeated endogenous matching game and also show that it has some very useful properties. 

\subsubsection{Equilibrium analysis for the repeated endogenous matching game.} 

For our mechanism $\boldsymbol{m}^F$, the action of the workers consists of the effort to exert in the assessment phase and the operational phase, while in the reporting phase the actions for the workers consists of both the effort to exert and the preference lists to report. 
Next, we propose a strategy for each worker $i$, which we refer to as MTBB (M-maximum, T-truthful, BB-bang-bang) strategy $\boldsymbol{\pi}_{i}^{MTBB}$ for the following reason.  A worker following MTBB exerts maximum effort in the assessment phase, then reports the preferences truthfully in the reporting phase, and then uses a bang-bang type structure for exerting effort (maximum or no effort) in the operational phase.
We will show that the MTBB strategy maximizes the long-run utility of the worker.

\begin{compactenum}
	\item \textbf{Assessment phase} ($0 \leq t\leq N-1$)  In each time slot $t$ in this phase, where $t\leq N$,  worker $i$ should exert the maximum effort possible, given as $e_{i\boldsymbol{m}^{F}(\boldsymbol{h}_{0}^{t})[i]}^{max}$, where $\boldsymbol{m}^{F}(\boldsymbol{h}_{0}^{t})[i]=(t+i) \mod N$. In each time slot, the worker receives a payment from the matched client and also observes the cost for exerting effort. We denote the payment received by worker $i$ in time slot $t$ as $\boldsymbol{P}(i, \boldsymbol{m}^{F}(\boldsymbol{h}_{0}^{t})[i])$ and the cost incurred by worker $i$ in time slot $t$ as $\boldsymbol{\bar{C}}(i, \boldsymbol{m}^{F}(\boldsymbol{h}_{0}^{t})[i])$. At the end of this phase,  worker $i$ knows the $\boldsymbol{P}(i,x)$ and $\boldsymbol{\bar{C}}(i,x)$ for all the tasks $x\in \mathcal{S}$. 
	\item \textbf{Reporting phase} ($t=N$) The  worker $i$ constructs the vector of  long-run utilities that the worker expects to derive by being matched as follows, $\boldsymbol{U}(i,x)=\boldsymbol{P}(i,x)-\boldsymbol{\bar{C}}(i,x), \;\forall x\in \mathcal{S}$. The worker submits a truthful ranking, which is the ranking in the decreasing order of $\boldsymbol{U}(i,x)$. Worker $i$ exerts maximum effort on  task   assigned to it in this time slot.
	\item \textbf{Operational phase ($t\geq N+1$)} The planner executes the G-S algorithm (as described above) and assigns to worker $i$ a task $y$.   If $ \boldsymbol{U}(i,y)>0 $, then the worker exerts maximum effort $e_{iy}^{max}$ in every time slot, and otherwise the worker exerts zero effort in every time slot. 
	
\end{compactenum}

%

In the next theorem, we show that the proposed MTBB is a weakly dominant  strategy for each worker. Therefore, if all the workers follow the MTBB strategy, then the joint strategy will comprise an equilibrium of the repeated endogenous matching game (induced by the proposed mechanism $\Omega^F$), which we refer to as the bang-bang equilibrium (BBE).


\begin{theorem}\label{theorem1}\textbf{MTBB strategy and its properties}
	\begin{compactenum} 
		\item The MTBB strategy is a weakly dominant strategy for each worker.
		\item If all the workers follow the MTBB strategy, then the joint strategy is an  equilibrium.
	\end{compactenum}
\end{theorem}
%
See Appendix  for the proof of Theorem \ref{theorem1}.

The MTBB strategy is only weakly dominant and thus it does not imply that the bang-bang equilibrium is the unique NE.  In order to play MTBB, the worker does not need information about the strategy of other workers. We study the uniqueness of the BBE in the next section. \footnote{ If we consider the case where the workers also have some knowledge of the form of the distribution of the productivities of other workers, then as well the above theorem continues to hold because the MTBB strategy is a dominant strategy. Therefore, the bang-bang equilibrium will be a Bayesian Nash equilibrium.}

\textbf{Mechanism incentivizes truthful revelation and no hindrance in learning.}   The structure of the mechanism ensures that if a worker exerts maximum effort on one task, then there is no decrease in the chance of getting accepted by a task that the worker prefers more.  Our design involves reporting of preferences from the worker side. Using  the G-S algorithm with workers as proposers incentivizes truthful revelation \cite{roth1982economics} and gives workers no incentive to hinder learning through their actions. In mechanisms that only operate based on the output and try to achieve efficient long-run performance, it can be shown that workers can strategically try to under-perform on some tasks. 


Before we analyze properties of the BBE we state certain assumptions  and we will invoke them in theorems as appropriate.

\textbf{Assumptions.} 

\begin{assumption}
	\textbf{ Better productivity means lower costs.}  $\boldsymbol{F}(i,x)>\boldsymbol{F}(k,x) \iff \boldsymbol{C}(k,x)>\boldsymbol{C}(i,x) \iff e_{ix}^{max} > e_{kx}^{max},\; \forall i,k \in \mathcal{N}, \;\forall x \in \mathcal{S} $
\end{assumption}

Assumption 1 states that if worker $i$ has a higher productivity  than another worker $k$ on a task $x$, then it has a lower cost for exerting effort on the same task and this is true for all the tasks $x\in \mathcal{S}$ and vice-versa. The same condition holds for the maximum effort. 
Assumption 1 is natural in  many settings. It states that if a worker has more experience (and skill) in performing a task, then the worker also has more interest in that task and is willing to spend more time on it.

\begin{assumption}
	\textbf{Task homogeneity.} The productivity, the cost of exerting effort, and the maximum effort of a worker,  is the same across all the tasks, i.e. $\boldsymbol{F}(i,x)=\boldsymbol{F}(i,y), \; \forall x, y$ and is denoted as $F(i)$,      
	$\boldsymbol{C}(i,x)=\boldsymbol{C}(i,y), \;\forall x,y$ and is denoted as $C(i)$, $e_{ix}^{max}=e_{iy}^{max}, \;\forall x,y$ and is denoted as $e_{i}^{max}$.
\end{assumption}

Assumption 2 states that the type of a worker across the different tasks are the same. This is natural in settings where the tasks are homogeneous (of the same type).  For instance, all the tasks can relate to a particular language of software development. This assumption requires homogeneity in task types but still allows the tasks to have different qualities. For instance, different software development tasks can generate different revenues (qualities). Moreover, different workers can still have different qualities over the tasks even though the tasks are of the same type.

\begin{assumption}
	\textbf{Task distribution structure.} The quality of a task $g(x)$ is either more than $g_u$ (high quality task) or less than $g_l$ (low quality task). 
\end{assumption}

The above assumption ensures that if a task's quality is greater than $g_u$ (expression in the Appendix), then every worker wants to exert non-zero effort on it, else if the task's quality is lower than $g_l$ (expression in the Appendix), then no worker wants to exert effort on it. This assumption can also be interpreted as the task distribution having at least two modes (bimodal).  Next, we analyze properties of the BBE.


\begin{theorem} \label{theorem2} \textbf{Long-run stability.}  If Assumption 1 holds, then the  bang-bang equilibrium is long-run coalition-stable under FILI matching mechanism.
\end{theorem}

See Appendix for the proof of Theorem \ref{theorem2}. 


The above Theorem also implies that the BBE is long-run pairwise stable. Next, we compare the pairwise stability aspect with existing results in the literature. Our Theorem \ref{theorem2} bears similarity to Theorem 5 in \cite{roth1982economics}. In Theorem 5 in \cite{roth1982economics} it is shown that if the matching rule is worker-optimal and outputs stable outcomes (stability in the sense of \cite{gale1962college}), then the truthful revelation of preferences is the dominant strategy for all the workers. Recall that in our setting, the preference list submitted in the MTBB strategy corresponds to the true preference list.
In Theorem \ref{theorem2}, we prove that if the proposed mechanism is used (it uses worker-optimal matching in the operational phase), then we know that for every worker MTBB strategy, which leads to the truthful revelation of preferences, is a dominant strategy and is long-run stable. In both \cite{roth1982economics} and our setting, it is shown that it is possible to achieve truthful revelation on the worker side and also achieve stability.




\textbf{Uniqueness of the equilibrium.}  In the next theorem, we show that in many cases the repeated endogenous matching game has a unique equilibrium payoff (vector of long-run utilities of the workers), which is achieved by the bang-bang equilibrium strategy. 
Note that the uniqueness in terms of payoffs means that there can be multiple equilibrium strategies possible but all of them lead to the same unique equilibrium payoff.

\begin{theorem} \label{theorem3} \textbf{Uniqueness of the equilibrium payoff.} 
	If the Assumptions 1, and 2 hold, then the repeated endogenous matching game induced by the FILI matching mechanism has a unique equilibrium payoff, which is achieved by the bang-bang equilibrium strategy.
\end{theorem}

See Appendix for the proof of Theorem \ref{theorem3}.
Next, we  establish the conditions under which the proposed mechanism is effective in mitigating the moral hazard and enables effective learning on both sides and thus achieves the optimal long-run revenue.

\begin{theorem} 
	\label{theorem4}
	\textbf{Optimal long-run revenue.} 
	If Assumptions 2 and 3 hold, then the FILI matching mechanism $\boldsymbol{m}^F$ achieves the optimal total long-run revenue among all the mechanisms $\mathcal{M}$.
\end{theorem}

See Appendix for the proof of Theorem \ref{theorem4}.

\section{Numerical Experiments}






In this section, we present numerical experiments to show that the performance achieved by the proposed mechanism is very high.  
We assume that each worker's productivity, the cost for exerting effort on every task, and the task qualities is drawn from distributions that are known to the planner. Further details of the setup for the numerical experiments can be found in the Appendix.  

Note that the distribution of tasks does not follow the Assumption 3 that was used to prove Theorem \ref{theorem4}. We also show that our mechanism is not restricted to quadratic payments and the results presented extend to different payment rules. The set of payment rules we use here is the union of the following two families-
i) Linear payments: A worker is paid a fixed amount per unit output that it generates. Specifically, a worker is paid a fraction of the revenue generated, where the fraction is a parameter of the payment rule that needs to be selected by the planner.
ii) Quadratic payments:  The client $x$ pays a worker $\alpha w^2 g(x)$ amount for producing $w$ units of output. 

Next, we describe the mechanisms that we compare.

1. \textbf{Initial belief based matching combined with optimal payment.}  The planner matches the workers based on its initial beliefs about the workers as follows. The planner ranks the workers based on the mean of the beliefs across the tasks and matches the workers with the tasks assortatively, where the tasks are ranked based on their qualities. If two workers share the same rank, then the matching is done randomly for those workers. With this as the matching rule, the planner selects the optimal payment rule from the above family of  payment rules to optimize the chosen performance criterion, which can be the total long-run revenue or the total long-run profit. 

In many existing setups, the matching rules are similar to this initial belief based matching. For instance, on Upwork the clients and the workers are matched based on the initial information provided. Also, the firms that do not practice job rotation \cite{ortega2001job} follow a similar mechanism  that relies only on the initial beliefs. The payment rules on platforms such as Upwork generally follow a linear payment structure.


2. \textbf{Proposed mechanism combined with optimal payment.}  For our mechanism, we will use the proposed FILI matching rule $\boldsymbol{m}^F$. The planner selects the optimal payment rules from the same family of  payment rules described at the beginning of this section  (given the fixed choice of matching rule  $\boldsymbol{m}^F$).

3. \textbf{Upper bound on the total long-run revenue and the profit.} We use the  upper bound for the total long-run revenue and the total long-run profit that we derive in the Appendix. 
\begin{figure}[t]
	\begin{center} 
		\includegraphics[trim=0 3cm 0 0.5cm, width=4.8in]{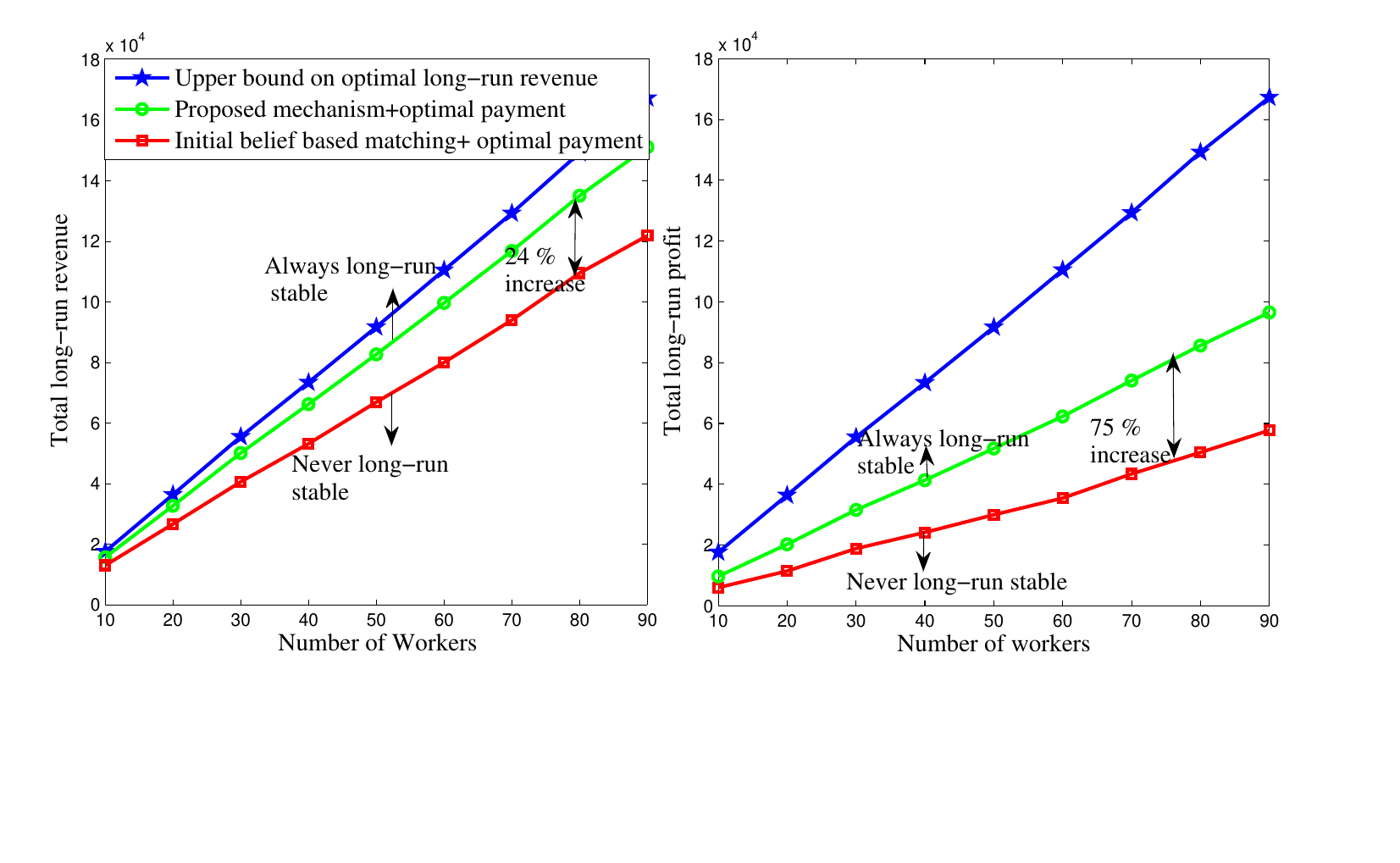}
		\caption{Comparison of the proposed mechanism with other approaches.}
		\label{comparison1}
		
	\end{center}
\end{figure}

%


%

In Figure \ref{comparison1}, we compare the performance of the proposed mechanism  combined with optimal payment against the mechanisms described above and the upper bound derived in  the Appendix.  It can be seen that the proposed mechanism leads to large gains of over 75 percent and is always long-run stable.


\section{Extensions}
\label{extsec}

\textbf{Payment rules decided by the clients.} In Section \ref{sec2}, we considered the settings where the payment rules were given and the clients were required to comply. We can extend some of the important results presented in this work to settings where the choice of payment rules is a part of the client's strategy and are not set by the planner.   For ease of exposition, we will assume that each client has to choose from a set of linear payment rules- client pays the worker a fraction of the revenue generated, where the fraction is decided by the client. The costs for exerting effort for the workers is a linear function in the exerted effort as well. Suppose that Assumptions 1 and 2 hold.  We also assume that the clients know the distribution from which the workers are drawn and vice-versa.
Under these conditions, we can arrive at an equilibrium strategy, which is very similar to the bang-bang equilibrium strategy. We can also show that the matching achieved is long-run stable with respect to this equilibrium strategy. The only new component in the equilibrium strategy that needs explanation are clients' payment rules. The client with the highest quality task will want to attract the worker with highest quality (since Assumptions 1 and 2 hold).  The client with highest task quality will need to use a payment rule that guarantees that the worker with highest quality is paid at least as much as being offered by the client with second highest task quality. The same argument applies to the client with second highest task quality and so on. In such a case, all the clients will set  payment rules such that the amount paid per unit output to all the workers  is the same. Further details are in the Appendix.

We also discuss some other extensions in Appendix.

%

\section{Conclusion}

In this work, we consider an environment with career concerns, where the workers are assessed by different clients over time before finally getting matched to a particular client. The mechanism considered  requires the planner to take actions based on the  outputs produced by the self-interested workers, where the outputs depend on both productivity of workers  and efforts exerted (unobserved thus moral hazard).  We consider the settings where the workers and clients are extremely patient and want to optimize the long-run utilities, which are defined as the limit of means of stage game utilities. Hence, any experimentation carried out for a finite time to assess the matches is costless.
We construct a mechanism that achieves optimal total long-run revenue in the equilibrium when the tasks are  homogeneous (the worker's qualities across tasks do not vary) and have a bimodal distribution. 
We propose a notion of stability - ``long-run stability'', which is  meaningful for matching environments with incomplete information and learning. In a wide-range of settings, we prove that our proposed mechanism achieves  long-run stability. 


\section{Appendix}

%
\begin{table}

	\begin{center}
			\label{table1}
			\caption{Comparison of works in the area of matching, AS: Adverse Selection, MH: Moral Hazard}
		\begin{tabular}{| l | l  |  l | l | l| }
			\hline
			&	Matching  & Strategic    & Incomplete & MH, AS  \\ 
			&	 with transfer & workers & information &  \\  \hline
			\cite{gale1962college}	&	No & No  & No  & No, No\\ \hline
			\cite{shapley1971assignment}	& 	Yes & No & No & No, No \\	\hline
			\cite{shimer2000assortative}, & Yes & Yes & No & No, No  \\ 
			\cite{becker1974theory},& & & &\\ 
			\cite{grossman2013matching} & & & &\\ \hline
			\cite{roth1982economics}, & No & Yes & No & No, No\\
			\cite{immorlica2015incentives}  & & & & \\
			\cite{fragiadakis2016strategyproof}  & & & & \\
			\hline
			\cite{roth1989two},  \cite{liu2014stable} & No & Yes  & Yes (no learning) & No, Yes \\ 
			\cite{bikhchandani2014two} & & & &\\
			\hline		
			\cite{rastegari2013two},  & No  & No  & Yes (with learning) & No, Yes\\ 
			\cite{lee2009interviewing} & & & & \\ \hline
			
			\cite{hopkins2012job} & Yes & Yes & Yes (with learning) & No, Yes \\ \hline
			\cite{tran2014efficient} & No  & No & Yes (with learning) & No, Yes \\ \hline
			
			\cite{ho2012towards} & Yes & Yes & Yes (no learning) & Yes, No \\ \hline
			
			\cite{lazarova2017paths} & No & Yes & Yes (with learning) & No, Yes  \\ \hline
			\cite{ashlagi2015unbalanced}	& 	No & Yes & Yes (with learning) & No, Yes \\
			\hline
			\cite{arnosti2014managing}	& 	Yes & Yes & Yes (with learning) & No, Yes \\
			\hline
			\cite{kocer2014dynamic} & No & Yes& Yes (with learning) & No, Yes \\ \hline
			\cite{karger2014budget} & Yes & No & Yes (with learning)   & No, Yes \\ \hline
			\cite{dayama2015truthful} & Yes & Yes & Yes (no learning ) & No, No\\ \hline
			\cite{fisher2015matching} & Yes & Yes & No & Yes, No \\ \hline
			
			\cite{corbae2003directed} & Yes & Yes & No & No, No \\ \hline
			\cite{xiao2014incentive} &  Yes &  Bounded     & Yes (learning & \\ 
			& &  rational&  out of equilibrium)  & Yes, Yes \\   \hline
			\cite{xiao2016} & Yes &  Yes  & Yes (no learning) & Yes, Yes \\ \hline
			This work & Yes & Yes & Yes (with learning) & Yes, Yes \\ \hline
		\end{tabular}
	
	\end{center}
\end{table}

\begin{table}

	\begin{center}
				\caption{Knowledge structure summary}
			\label{table2}
		\begin{tabular}{| l  l    l  l | }
			\hline
			&	Client & Worker & Planner  \\ \hline
			Matching rule	& 		\checkmark &	\checkmark & \checkmark \\	\hline
			Payment rule &\checkmark & &   \checkmark\\ \hline
			Cost  $\boldsymbol{C}(i,x)$& & & \\ \hline 
			Productivity $\boldsymbol{F}(i,x)$ & & & \\ \hline 
			Task quality $g(x)$ & & & \\ \hline 
			Effort  & & \checkmark & \\ \hline
			Output &\checkmark && \checkmark\\ \hline 
		\end{tabular}

	\end{center}
\end{table}

In all the proofs we will use $I(A)$ as the indicator function. If the condition $A$ holds, then the indicator is one and zero otherwise.

\subsection{Proof of Proposition \ref{prop1}} 
It is easy to see that the workers can always ensure a zero long-run utility (outside option of the worker gives zero utility) by exerting zero effort. Therefore, the participation constraint for the workers is trivially satisfied.  If $\alpha \leq \frac{1}{2W^{max}}$, then the profit per unit output is always greater than or equal to zero which ensures that the clients cannot have a negative profit in any period. Thus the clients cannot have a negative long-run profit.

\subsection{Proof of Theorem \ref{theorem1}} 

From Proposition \ref{prop1}, we know that the clients will participate in the mechanism. Hence, in this proof we only focus on the worker's strategies.
There are two parts to the Theorem. In the first part, we need to show that the MTBB  strategy is a weakly dominant strategy. First, we will simplify the expression for the long-run utility of the worker $i$ when the proposed mechanism $\boldsymbol{m}^F$ is implemented. We write the joint strategy for all the workers  as $\boldsymbol{\pi}=(\boldsymbol{\pi}_{1},...,\boldsymbol{\pi}_{N})$. In Section \ref{sec2}, where we defined the strategy of the workers for a given mechanism, we used a general definition for the action set. The strategy consisted of two parts, $\boldsymbol{\pi}_{i}(h_{i}^{t})[1]$ is the effort exerted by worker $i$ and $\boldsymbol{\pi}_{i}(h_{i}^{t})[2]$ is the reports submitted by the worker. For our proposed mechanism $\Omega^F$, the second component of reports only plays a role in time slot $N$, which is the reporting phase, and for the rest of the time slots the clients can choose to send no reports as it does not impact the interactions in any way.

We write the private history of worker $i$, which is induced by the joint strategy $\boldsymbol{\pi}$ as $h_{i}^{t,\boldsymbol{\pi}}$. We write the preference list provided by  worker $i$  in the reporting phase as
\begin{equation}
\boldsymbol{b}_i=\boldsymbol{\pi}_{i}(h_{i}^{N,\boldsymbol{\pi}})[2] 
\label{eqn-proof-thm1-preflist}
\end{equation}
The output produced in time slot $t$ by worker $i$ assigned to task $j=(t+i) \mod N$ is written as
\begin{equation} \boldsymbol{W}^{e}(i,j)=\boldsymbol{F}(i,j)\boldsymbol{\pi}_{i}(h_{i}^{t,\boldsymbol{\pi}})
\label{eqn-proof-thm1-output_matrix}
\end{equation}

The G-S algorithm executed by the planner at the  beginning of the operational phase takes as input the preference lists $\{\boldsymbol{b}_{i}\}_{i=1}^{N}$ and the outputs produced by the workers $\boldsymbol{W}^{e}$. We represent the output of the G-S algorithm as 
\begin{equation}
\boldsymbol{m}^{GS}(\{\boldsymbol{b}_{i}\}_{i=1}^{N}, \boldsymbol{W}^{e})
\label{eqn-proof-thm1-mgs}
\end{equation}
where $\boldsymbol{m}^{GS}$ is a function that takes the preference lists and performance of workers as input and outputs the matching.  The joint strategy $\boldsymbol{\pi}$ induces an observation history for the planner, which we write as $\boldsymbol{h}_{0}^{t,\boldsymbol{\pi}}$. Note that $\boldsymbol{h}_{0}^{t,\boldsymbol{\pi}}$ and $\{\{\boldsymbol{b}_{i}\}_{i=1}^{N}, \boldsymbol{W}^{e}\} $ contain the same relevant information needed for the final matching to be determined by G-S algorithm.  For consistency, we state that when $t\geq N+1$,
\begin{equation}
\boldsymbol{m}^F(\boldsymbol{h}_{0}^{t,\boldsymbol{\pi}})=\boldsymbol{m}^{GS}(\{\boldsymbol{b}_{i}\}_{i=1}^{N}, \boldsymbol{W}^{e})
\label{eqn-proof-thm1-match_fin}
\end{equation} 
is the notation for the proposed matching rule given in Section \ref{sec2}. 

The expression for the long-run utility for  worker $i$ defined in Section \ref{sec2} is simplified by substituting \eqref{eqn-proof-thm1-match_fin} as follows.

\begin{equation}
\begin{split}
& U_i(\{\boldsymbol{\pi}_k\}_{k=1}^{N}| \boldsymbol{m}^F) =\\ & \lim_{T\rightarrow \infty}\frac{1}{T+1}\sum_{t=N+1}^{T} \Big[\alpha \boldsymbol{F}(i,\boldsymbol{m}^{GS}(\{\boldsymbol{b}_k\}_{k=1}^{N}, \boldsymbol{W}^e)[i] )^2 g(\boldsymbol{m}^{GS}(\{\boldsymbol{b}_k\}_{k=1}^{N}, \boldsymbol{W}^e)[i]) -\\&
\boldsymbol{C}(i, \boldsymbol{m}^{GS}\left(\{\boldsymbol{b}_k\}_{k=1}^{N}, \boldsymbol{W}^e[i]\right) \Big] (e_i^{t})^2 \label{Utility-exp}
\end{split}
\end{equation}
In the above expression \eqref{Utility-exp},  $e_{i}^{t}=\boldsymbol{\pi}_{i}(h_{i}^{t,\boldsymbol{\pi}})[1]$. In the above expression \eqref{Utility-exp}, we did not write the utility from the assessment and reporting phase because the number of time slots in assessment phase are finite $N+1$ and thus utilities in the assessment phase do not contribute to the long-run utility.

We define 
\begin{equation} \bar{e_{i}^2}=\lim_{T\rightarrow \infty }\sum_{t=0}^{T}\frac{1}{T+1} (e_{i}^{t})^2  
\label{eqn-proof-thm1-eiav}
\end{equation} 
We define  \begin{equation}
\begin{split}
&\boldsymbol{H}_{i}(\{\boldsymbol{b}_k\}_{k=1}^{N}, \boldsymbol{W}^e)= \\ & \alpha \boldsymbol{F}(i,\boldsymbol{m}^{GS}(\{\boldsymbol{b}_k\}_{k=1}^{N}, \boldsymbol{W}^e)[i] )^2 g(\boldsymbol{m}^{GS}(\{\boldsymbol{b}_k\}_{k=1}^{N}, \boldsymbol{W}^e)[i]) -
\boldsymbol{C}(i, \boldsymbol{m}^{GS}\left(\{\boldsymbol{b}_k\}_{k=1}^{N}, \boldsymbol{W}^e[i]\right)
\end{split}
\label{eqn-proof-succinct}
\end{equation}

Thus we can simplify the above utility \eqref{Utility-exp} by substituting \eqref{eqn-proof-thm1-eiav}, \eqref{eqn-proof-succinct} as follows. 
\begin{equation}U_i(\{\boldsymbol{\pi}_k\}_{k=1}^{N}| \boldsymbol{m})  = \bar{e_{i}^2}\boldsymbol{H}_{i}(\{\boldsymbol{b}_k\}_{k=1}^{N}, \boldsymbol{W}^e) \label{max-longrun}\end{equation}

Next, we want to solve for the optimal strategy $\boldsymbol{\pi}_{i}$ given the fixed strategy of the rest of the workers $\boldsymbol{\pi}_{-i}$. Formally stated, the optimization problem is given as follows.

\begin{equation}
\max_{\boldsymbol{\pi}_i} U_{i}(\{\boldsymbol{\pi}_{k}\}_{k=1}^{N}|\boldsymbol{m}^F)
\label{eqn-proof-thm1-maxim-utility}
\end{equation}

We will first compute an upper bound for \eqref{max-longrun}.  
Observe that
\begin{equation}
\begin{split}
& U_i(\{\boldsymbol{\pi}_k\}_{k=1}^{N}| \boldsymbol{m})  =\\& \bar{e_{i}^2}\boldsymbol{H}_{i}(\{\boldsymbol{b}_k\}_{k=1}^{N}, \boldsymbol{W}^e) \leq  (e_{i\boldsymbol{m}^{GS}(\{\boldsymbol{b}_k\}_{k=1}^{N}, \boldsymbol{W}^e)[i]}^{max})^2\boldsymbol{H}_{i}(\{\boldsymbol{b}_k\}_{k=1}^{N}, \boldsymbol{W}^e)  I\Big(\boldsymbol{H}_{i}(\{\boldsymbol{b}_k\}_{k=1}^{N}, \boldsymbol{W}^e)\geq 0 \Big)
\label{UB-utility-1}
\end{split}
\end{equation}

In the above expression \eqref{UB-utility-1}, the LHS will achieve the same value as the RHS provided worker $i$ follows the following strategy. If  $t\geq N+1$ and $\boldsymbol{H}_{i}(\{\boldsymbol{b}_k\}_{k=1}^{N}, \boldsymbol{W}^e)\geq 0$, then $e_{i}^{t}=e_{i\boldsymbol{m}^{GS}(\{\boldsymbol{b}_k\}_{k=1}^{N}, \boldsymbol{W}^e)[i]}^{max}$ and $e_{i}^{t}=0$ zero otherwise.  We now compute the optimal value for the maximum for the RHS. The expression in RHS depends only on the actions taken in the assessment and the reporting phase. 
Based on the above inequality \eqref{UB-utility-1}, we can say that the optimizer of the RHS in terms of the actions in the assessment and reporting phase will be an upper bound of the maximization problem in \eqref{max-longrun}. 
We first maximize the expression in RHS with respect to the choice of preference lists submitted in the reporting phase. 

We  claim that if worker $i$ ranks the clients in the order of $ [\alpha \boldsymbol{F}(i,j)^2g(j)- \boldsymbol{C}(i,j)](e_{ij}^{max})^2$ for all $j$, then it corresponds to the best choice of the preference list. We denote this preference list as $\boldsymbol{b}_{i}^{*}$. This claim follows from Theorem 5 \cite{roth1982economics}, where it is shown that the truthful reporting is a dominant strategy when the matching rule is worker optimal and leads to stable outcomes. 

Thus we can write

\begin{equation}
\begin{split}
U_i(\{\boldsymbol{\pi}_k\}_{k=1}^{N}| \boldsymbol{m})  = \\
\bar{e_{i}^2}\boldsymbol{H}_{i}(\{\boldsymbol{b}_k\}_{k=1}^{N}, \boldsymbol{W}^e)
\leq (e_{i\boldsymbol{m}^{GS}(\{\boldsymbol{b}_k\}_{k=1}^{N}, \boldsymbol{W}^e)[i]}^{max})^2\boldsymbol{H}_{i}(\{\boldsymbol{b}_k\}_{k=1}^{N}, \boldsymbol{W}^e)I\Big(\boldsymbol{H}_{i}(\{\boldsymbol{b}_k\}_{k=1}^{N}, \boldsymbol{W}^e)\geq 0 \Big) \leq \\
(e_{i\boldsymbol{m}^{GS}(\{\boldsymbol{b}_k\}_{k=1, k\not=i }^{N}, \boldsymbol{b}_{i}^{*}, \boldsymbol{W}^e)[i]}^{max})^2\boldsymbol{H}_{i}(\boldsymbol{b}_{i}^{*}, \{\boldsymbol{b}_k\}_{k=1, k\not=i }^{N}, \boldsymbol{W}^e)I\Big(\boldsymbol{H}_{i}(\boldsymbol{b}_{i}^{*}, \{\boldsymbol{b}_k\}_{k=1, k\not=i }^{N}, \boldsymbol{W}^e) \geq 0 \Big)
\label{UB-Utility}
\end{split}
\end{equation}
Next, we will show that if the preference list is fixed for worker $i$ to $\boldsymbol{b}_{i}^{*}$, then the choice of effort level for task $j$ in the assessment phase, which is denoted as $e_{ij}^{eval}$, that maximizes the RHS of the above expression \eqref{UB-Utility} is $e_{ij}^{max}$. We do so by arguing that the long-run utility of the worker increases in $e_{ij}^{eval}$. 

If the worker increases $e_{ij}^{eval}$ to $e_{ij}^{eval}+\delta$, then the ranking of the worker by  task $j$ can either stay the same or increase. Since other parameters remain the same, the ranking of  worker $i$ on other tasks does not change. In this case, there are three possibilities. Suppose that the worker exerts effort levels $\{e_{ik}^{eval}\}_{k=1}^{N}$ in the assessment phase on different tasks and submits the preference list $\boldsymbol{b}_{i}^{*}$ in the reporting phase and is matched to task $j_1$ in the operation phase. We will analyze the behavior of the \eqref{UB-Utility} when we vary the effort level of worker $i$  on task $j$ $e_{ij}^{eval}$. It is possible that rank of task $j_1$ in the preference list $\boldsymbol{b}_{i}^{*}$ is greater than task $j$ or equal or lesser. If the rank of $j_1$ is greater than $j$, then the worker even after increasing effort on task $j$ will still be accepted by $j_1$ as the ranking of the worker for $j_1$ and ranking of $j_1$ for all workers is not affected by $e_{ij}^{eval}$. Thus, in this case, increasing the effort $e_{ij}^{eval}$ will not change the rank of the task that is assigned.

If the rank of $j_1$ is equal to $j$, then by increasing the effort $e_{ij}^{eval}$ can only improve worker's ranking for task $j$. The ranking of worker $i$ on tasks ranked higher than task $j$ is still the same, thus worker $i$ will be rejected by all those tasks. But since the ranking of worker $i$ on task $j$ is the same or higher it means that the worker will be assigned to $j$. 

If the ranking of task $j_1$ is lesser than the rank of task $j$, then note that the ranking of the worker on task $j_1$ will not change and thus the worker will still be accepted by task $j_1$ at least. However, since the worker increases effort on task $j$ the ranking of the worker can improve on task $j$. This means that it is possible that the worker is accepted by a strictly higher ranked task. Thus we know that increasing effort $e_{ij}^{eval}$ can lead to the worker being matched to a task with higher or the same rank as before. A task with higher or the same rank will imply a higher or the same value for the long-run utility of the worker. Hence, the $e_{ij}^{eval}=e_{ij}^{max}$ is the optimal choice at which the upper bound in the RHS is maximized. This holds for all the tasks that  worker $i$ is matched to for the first time in the assessment phase.
Observe that the proposed MTBB strategy achieves the value for the upper bound in the RHS, thus it has to be the best response for a worker to every strategy of other workers.

The next part of the theorem follows easily from the fact that since all the workers use their best response strategies the joint strategy has to be an equilibrium.

\subsection{Proof of Theorem \ref{theorem2}}

We first prove long-run pairwise stability. 
Before we give the Proof for Theorem \ref{theorem2}, we first need to simplify and arrive at the expressions for the long-run utilities for the workers and clients as given in Section \ref{sec2}. We only consider the matching rules for which the limit of the matching exists across all the histories.  Suppose the joint strategy being used by the workers and the clients is $\boldsymbol{\pi}$. Under this joint strategy the limit of the matching rule is given as $\boldsymbol{m}^{*}_{\boldsymbol{\pi}}$. The history that is induced by the joint strategy $\boldsymbol{\pi}$ is defined as $\boldsymbol{h}_{0}^{t, \boldsymbol{\pi}}$ and the $\boldsymbol{h}_{i}^{t, \boldsymbol{\pi}}$ for worker $i$. Note that the $\lim_{t\rightarrow \infty}\boldsymbol{m}(\boldsymbol{h}_{0}^{t, \boldsymbol{\pi}})=\boldsymbol{m}^{*}_{\boldsymbol{\pi}}$, where the limit is defined using the standard Euclidean norm in the space $\mathbb{R}^{N}$ as the distance metric. Next, we will show that the above limit is attained after a finite number of time slots denoted as $T_{lim}$. Note that the minimum distance between any two distinct matchings is finite and is given as $d_{min}$. From the definition of limit, it is clear that there exists a constant $T_{lim}$ such that if $t\geq T_{lim}$, then the distance between $\boldsymbol{m}(\boldsymbol{h}_{0}^{t, \boldsymbol{\pi}})$ and $\boldsymbol{m}^{*}_{\boldsymbol{\pi}}$ is less than $d_{min}$. Therefore, for all $t\geq T_{lim}$ \begin{equation}\boldsymbol{m}(\boldsymbol{h}_{0}^{t, \boldsymbol{\pi}}) = \boldsymbol{m}^{*}_{\boldsymbol{\pi}}
\end{equation}

Based on the above simplification we can write the long-run utility of a worker $i$ and client $x=\boldsymbol{m}_{\boldsymbol{\pi}}^{*}[i]$ as follows.

\begin{equation}
\begin{split}
U_{i}(\{\boldsymbol{\pi}_{k}\}_{k=1}^{2N}| \boldsymbol{m}) &\textsl{} = \lim_{T\rightarrow \infty} \sum_{t=T_{lim}}^{T}\frac{1}{T+1} p(\boldsymbol{F}(i,x)\boldsymbol{\pi}_{i}(\boldsymbol{h}_{i}^{t, \boldsymbol{\pi}}), x)-\boldsymbol{C}(i,x) \boldsymbol{\pi}_{i}(\boldsymbol{h}_{i}^{t, \boldsymbol{\pi}})^2 \\ 
& = \lim_{T\rightarrow \infty} \sum_{t=0}^{T}\frac{1}{T+1} p(\boldsymbol{F}(i,x)\boldsymbol{\pi}_{i}(\boldsymbol{h}_{i}^{t, \boldsymbol{\pi}}), x)-\boldsymbol{C}(i,x) \boldsymbol{\pi}_{i}(\boldsymbol{h}_{i}^{t, \boldsymbol{\pi}})^2 
\end{split}
\end{equation}

Similar justification applies for the clients' long-run utilities as well.

We write the matching achieved in the bang-bang equilibrium at the start of the operational phase as $\boldsymbol{m}^{BBE}$.  The long-run utility for worker $i$ in the bang-bang equilibrium $U_{i}(\{\boldsymbol{\pi}_{k}^{MTBB}\}_{k=1}^{N}| \boldsymbol{m}^F)$is simplified below.

\begin{equation}
\begin{split}
&\left[\alpha \boldsymbol{F}(i, \boldsymbol{m}^{BBE}[i])^2 g( \boldsymbol{m}^{BBE}[i])-\boldsymbol{C}(i,\boldsymbol{m}^{BBE}[i])\right](e_{i\boldsymbol{m}^{BBE}[i]}^{max})^2 \times \\ &I\Big(\alpha \boldsymbol{F}(i, \boldsymbol{m}^{BBE}[i])^2 g( \boldsymbol{m}^{BBE}[i])-\boldsymbol{C}(i,\boldsymbol{m}^{BBE}[i]) \geq 0\Big)
\end{split} \label{eqn-proof-thm4-utility-worker}
\end{equation}

Define $\boldsymbol{J}:\mathcal{N}\times \mathcal{S}\rightarrow \mathbb{R}$ and $\boldsymbol{L}:\mathcal{N}\times \mathcal{S}\rightarrow \mathbb{R}$ as follows.
\begin{equation}
\begin{split}
\boldsymbol{J}(k,x) = \left[\alpha \boldsymbol{F}(k, x)^2 g( x)-\boldsymbol{C}(k,x)\right] \\
\boldsymbol{L}(k,x) = I\Big(\alpha \boldsymbol{F}(k,x)^2 g( x)-\boldsymbol{C}(k,x) \geq 0\Big) \label{eqn-proof-thm4-utility-trm}
\end{split}
\end{equation}	
We can write \eqref{eqn-proof-thm4-utility-worker} using \eqref{eqn-proof-thm4-utility-trm} more succinctly as follows.
\begin{equation}
U_{i}(\{\boldsymbol{\pi}_{k}^{MTBB}\}_{k=1}^{N}| \boldsymbol{m}^F) =\boldsymbol{J}(i,\boldsymbol{m}^{BBE}[i])\boldsymbol{L}(i,\boldsymbol{m}^{BBE}[i])(e_{i\boldsymbol{m}^{BBE}[i]}^{max})^2  \label{eqn-proof-thm4-utility-worker-1}
\end{equation}
The long-run utility for  client $\boldsymbol{m}^{BBE}[m]$, where $m\not=i$, in the bang-bang equilibrium is given as follows.

\begin{equation}
\begin{split}
(1-\alpha \boldsymbol{F}(m, \boldsymbol{m}^{BBE}[m])e_{m\boldsymbol{m}^{BBE}[m]}^{max}) \\
I\bigg(\alpha \boldsymbol{F}(m, \boldsymbol{m}^{BBE}[m])^2g( \boldsymbol{m}^{BBE}[m]) - \boldsymbol{C}(m,\boldsymbol{m}^{BBE}[m] ) \geq 0\bigg)\boldsymbol{F}(m, \boldsymbol{m}^{BBE}[m])e_{m\boldsymbol{m}^{BBE}[m]}^{max}
\end{split}
\label{eqn-proof-thm4-utility-client}
\end{equation}

We can simplify \eqref{eqn-proof-thm4-utility-client} using \eqref{eqn-proof-thm4-utility-trm} as follows.
\begin{equation}
\begin{split}
(1-\alpha \boldsymbol{F}(m, \boldsymbol{m}^{BBE}[m])e_{m\boldsymbol{m}^{BBE}[m]}^{max}) \boldsymbol{L}(m,\boldsymbol{m}^{BBE}[m])\boldsymbol{F}(m, \boldsymbol{m}^{BBE}[m])e_{m\boldsymbol{m}^{BBE}[m]}^{max}
\end{split}
\label{eqn-proof-thm4-utility-client-1}
\end{equation}
Suppose worker $i$ is matched to client $\boldsymbol{m}^{BBE}[m]$ instead in the operational phase. Our objective here is to show that it is not possible for both worker $i$ and client $\boldsymbol{m}^{BBE}[m]$ to increase their long-run utilities by being matched to one another and this holds true for every $i\not=m$.

If the utility for  worker $i$ strictly increases by being matched to $\boldsymbol{m}^{BBE}[m]$, then it has to hold true that $\big[\alpha F(i, \boldsymbol{m}^{BBE}[m])^2g( \boldsymbol{m}^{BBE}[m])-\boldsymbol{C}(i, \boldsymbol{m}^{BBE}[m])\big](e_{i\boldsymbol{m}^{BBE}[m]}^{max})^2$ has to  be  strictly higher than $\big[\alpha \boldsymbol{F}(i, \boldsymbol{m}^{BBE}[i])^2g(\boldsymbol{m}^{BBE}[i])-\boldsymbol{C}(i,\boldsymbol{m}^{BBE}[i])\big](e_{i\boldsymbol{m}^{BBE}[i]}^{max})^2$. This has to hold true because otherwise  the maximum utility that worker $i$ can  achieve by getting matched to $\boldsymbol{m}^{BBE}[m]$  will always be lesser than or equal to the long-run utility that the worker can achieve by getting matched to task $\boldsymbol{m}^{BBE}[i]$ in the operational phase of the bang-bang equilibrium.

We can write the utility for worker $i$ when it is matched to $\boldsymbol{m}^{BBE}[m]$  in the operational phase and when it follows  strategy $\boldsymbol{\pi}_{i}^{'}$ as follows. As explained in Section \ref{sec2},  it is sufficient to consider the strategies $\boldsymbol{\pi}_{i}^{'}$ that only depend on time. Also, we want to clarify that  the  deviation from a client side $ \boldsymbol{\pi}_{\boldsymbol{m}^{BBE}[m]}^{'}$  means that the client would not participate in the mechanism. For the worker $i$ as well we would assume that when strategy $\boldsymbol{\pi}_{i}^{'}$  is followed the worker does not participate in the mechanism. The worker $i$ and client $\boldsymbol{m}^{BBE}[m]$ interact outside the platform. We now write the utilities of both the worker $i$ and client $\boldsymbol{m}^{BBE}[m]$.
\begin{equation}
\begin{split}
\hat{U}_{i}(\boldsymbol{\pi}_{i}^{'}, \boldsymbol{\pi}_{\boldsymbol{m}^{BBE}[m]}^{'}) =\boldsymbol{J}(i,\boldsymbol{m}^{BBE}[m]) \boldsymbol{L}(i,\boldsymbol{m}^{BBE}[m]) \lim_{T\rightarrow \infty}\sum_{t=N+1}^{T} \frac{\boldsymbol{\pi}_{i}^{'}(t)^2}{T+1} 
\label{eqn-proof-thm4-dev-utility-worker}
\end{split}
\end{equation}
We write $\lim_{T\rightarrow \infty}\sum_{t=N+1}^{T} \frac{\boldsymbol{\pi}_{i}^{'}(t)^2}{T+1}=\bar{e_{i}^{2}}$, $\lim_{T\rightarrow \infty}\sum_{t=N+1}^{T} \frac{\boldsymbol{\pi}_{i}^{'}(t)}{T+1}=\bar{e_{i}}$ and substitute in \eqref{eqn-proof-thm4-dev-utility-worker} to obtain the following.
\begin{equation}
\hat{U}_{i}(\boldsymbol{\pi}_{i}^{'}, \boldsymbol{\pi}_{\boldsymbol{m}^{BBE}[m]}^{'}) =\boldsymbol{J}(i,\boldsymbol{m}^{BBE}[m]) \boldsymbol{L}(i,\boldsymbol{m}^{BBE}[m])\bar{e_{i}^{2}} 
\end{equation}
Also, the utility for  client $\boldsymbol{m}^{BBE}[m]$ in this case is derived as follows. 

\begin{equation}
\begin{split}
&\hat{V}_{\boldsymbol{m}^{BBE}[m]}(\boldsymbol{\pi}_{i}^{'}, \boldsymbol{\pi}_{\boldsymbol{m}^{BBE}[m]}^{'}) \\  &=\lim_{T\rightarrow \infty}\sum_{t=N+1}^{T}\frac{1}{T+1}\left[1-\alpha \boldsymbol{F}(i, \boldsymbol{m}^{BBE}[m])\boldsymbol{\pi}_{i}^{'}(t)\right] \boldsymbol{L}(i, \boldsymbol{m}^{BBE}[m]) \boldsymbol{F}(i, \boldsymbol{m}^{BBE}[m])\boldsymbol{\pi}_{i}^{'}(t) \\
&=\left[\boldsymbol{F}(i, \boldsymbol{m}^{BBE}[m])\bar{e_{i}}-\alpha \boldsymbol{F}(i, \boldsymbol{m}^{BBE}[m])^2\bar{e_{i}^2}\right] \boldsymbol{L}(i, \boldsymbol{m}^{BBE}[m])  \\
& \leq \left[\boldsymbol{F}(i, \boldsymbol{m}^{BBE}[m])\bar{e_{i}}-\alpha \boldsymbol{F}(i, \boldsymbol{m}^{BBE}[m])^2(\bar{e_{i}})^2\right]\boldsymbol{L}(i, \boldsymbol{m}^{BBE}[m])
\label{eqn-proof-thm4-clt-ubd}
\end{split}
\end{equation}

Based on the G-S algorithm and the fact that every worker uses MTBB strategy, we know that the rank of worker $m$ is higher than the rank of worker $i$ for task $\boldsymbol{m}^{BBE}[m]$.
\begin{equation}
\boldsymbol{F}(m,\boldsymbol{m}^{BBE}[m])e_{m\boldsymbol{m}^{BBE}[m] }^{max}> \boldsymbol{F}(i,\boldsymbol{m}^{BBE}[m])e_{i\boldsymbol{m}^{BBE}[m]}^{max}
\label{eqn-proof-thm4-ineq-work}
\end{equation}
From the above \eqref{eqn-proof-thm4-ineq-work}, either the productivity or the maximum effort has to be strictly higher. From Assumption 1, we know that if one of them is true, then the other is also true. In addition, we can say the following:

\begin{equation}
\boldsymbol{F}(m,\boldsymbol{m}^{BBE}[m])\geq F(i, \boldsymbol{m}^{BBE}[m]) \implies  \boldsymbol{C}(i, \boldsymbol{m}^{BBE}[m]) \geq \boldsymbol{C}(m, \boldsymbol{m}^{BBE}[m]) 
\label{eqn-proof-thm4-ineq-work-1}
\end{equation}

Based on the above \eqref{eqn-proof-thm4-ineq-work-1}, we can show the following.
\begin{equation}
\boldsymbol{J}(m, \boldsymbol{m}^{BBE}[m]) \geq 	\boldsymbol{J}(i, \boldsymbol{m}^{BBE}[m]) 
\label{eqn-proof-thm4-ineq1}
\end{equation}
\begin{equation}
\boldsymbol{L}(m, \boldsymbol{m}^{BBE}[m]) \geq 	\boldsymbol{L}(i, \boldsymbol{m}^{BBE}[m]) 
\label{eqn-proof-thm4-ineq2}
\end{equation}
Observe that the function $(1-\alpha x)x$ is increasing in $[0, \frac{1}{2\alpha}]$. We assumed that $\alpha \leq \frac{1}{2W^{max}}$. Therefore,  $(1-\alpha x)x$ is increasing is $x\in [0, W^{max}]$.
Note that \begin{equation}W^{max}\geq \boldsymbol{F}(m,\boldsymbol{m}^{BBE}[m])e_{m \boldsymbol{m}^{BBE}[m]}^{max}\geq \boldsymbol{F}(i,\boldsymbol{m}^{BBE}[m])e_{i\boldsymbol{m}^{BBE}[m]}^{max} \geq  \boldsymbol{F}(i,\boldsymbol{m}^{BBE}[m])\boldsymbol{\pi}_{i}^{'}(t) \label{eqn-proof-thm4-ineq-wm}
\end{equation}

We can use the above relations \eqref{eqn-proof-thm4-ineq-work-1},  \eqref{eqn-proof-thm4-ineq1}, \eqref{eqn-proof-thm4-ineq2}, \eqref{eqn-proof-thm4-ineq-wm} to derive the following condition on the expression in \eqref{eqn-proof-thm4-clt-ubd}.
\begin{equation}
\begin{split}
\left[\boldsymbol{F}(i, \boldsymbol{m}^{BBE}[m])\bar{e_{i}}-\alpha \boldsymbol{F}(i, \boldsymbol{m}^{BBE}[m])^2\bar{e_{i}}^2\right] \boldsymbol{L}(i, \boldsymbol{m}^{BBE}[m]) \leq \\
\left[\boldsymbol{F}(i, \boldsymbol{m}^{BBE}[m])e_{i\boldsymbol{m}^{BBE}[m]}^{max}-\alpha \boldsymbol{F}(i, \boldsymbol{m}^{BBE}[m])^2(e_{i\boldsymbol{m}^{BBE}[m]}^{max})^2\right]\boldsymbol{L}(i, \boldsymbol{m}^{BBE}[m])  \leq \\
\left[\boldsymbol{F}(m, \boldsymbol{m}^{BBE}[m])e_{m\boldsymbol{m}^{BBE}[m]}^{max}-\alpha \boldsymbol{F}(m, \boldsymbol{m}^{BBE}[m])^2(e_{m\boldsymbol{m}^{BBE}[m]}^{max})^2\right] \boldsymbol{L}(m,\boldsymbol{m}^{BBE}[m])
\end{split}
\label{eqn-proof-thm4-final-ineq}
\end{equation}

Therefore, from the above \eqref{eqn-proof-thm4-final-ineq}, we can see that the  client $\boldsymbol{m}^{BBE}[m]$ cannot have a strict gain at the same time as  worker $i$. Thus we can conclude that the proposed matching rule has to be long-run pairwise-stable w.r.t the bang-bang equilibrium strategy (joint MTBB strategy). We now move on to long-run coalition-stable. 

Let us assume that the BBE strategy is not long-run coalition stable. Therefore, we know that there exists a subset $\mathcal{X}$, which strictly benefits from deviating. We analyze three possibilities for the deviating subset. 
Suppose that the deviating subset consists of only workers. The joint strategy of the deviating subset and the equilibrium strategy of the non-deviating set outperforms the BBE strategy. 
Therefore, for each worker $i$  the following condition is true

$$ \hat{U}_{i}(\boldsymbol{\pi}_i^{'}, \boldsymbol{\pi}_{-i}^{'}, \boldsymbol{\pi}_{\mathcal{X}^{c}} ) > U_{i}(\boldsymbol{\boldsymbol{\pi}}^{BBE}|\boldsymbol{m}^F)$$
where $-i$ denotes the remaining indices of the individuals in the set $\mathcal{X}$, $ \boldsymbol{\pi}_{\mathcal{X}^{c}} $ is the set of strategies for individuals outside $\mathcal{X}$.
We know that $\boldsymbol{\boldsymbol{\pi}}_{i}^{BBE}$ is a weakly dominant strategy. Therefore,  $ \hat{U}_{i}(\boldsymbol{\pi}_i^{'}, \boldsymbol{\pi}_{-i}^{'}, \boldsymbol{\pi}_{\mathcal{X}^{c}} ) =  U_{i}(\boldsymbol{\pi}_i^{BBE}, \boldsymbol{\pi}_{-i}^{'}, \boldsymbol{\pi}_{\mathcal{X}^{c}} )$. 

Firstly, we know that $ U_{i}(\boldsymbol{\pi}_i^{'}, \boldsymbol{\pi}_{-i}^{'}, \boldsymbol{\pi}_{\mathcal{X}^{c}} ) >0$ otherwise it cannot be the case that 
$$ \hat{U}_{i}(\boldsymbol{\pi}_i^{'}, \boldsymbol{\pi}_{-i}^{'}, \boldsymbol{\pi}_{\mathcal{X}^{c}} ) > U_{i}(\boldsymbol{\boldsymbol{\pi}}^{BBE}|\boldsymbol{m}^F)$$
The minimum possible utility in the equilibrium is zero, i.e. $U_{i}(\boldsymbol{\boldsymbol{\pi}}^{BBE} | \boldsymbol{m}^F)>0$. 

We know that the utility that the worker $i$ gets is non-zero and 
based on our assumptions, we know that there are no ties in the preferences in the BBE. Therefore, by switching from $\boldsymbol{\pi}_{i}^{'}$ to $\boldsymbol{\pi}_{i}^{BBE}$ the worker $i$ is matched to the same task $z$ as it was matched under $\hat{U}_{i}(\boldsymbol{\pi}_i^{'}, \boldsymbol{\pi}_{-i}^{'}, \boldsymbol{\pi}_{\mathcal{X}^{c}} )$.

For a worker $j \not = i $ that is in $\mathcal{X}$, we calculate the impact when worker $i$ switches from $\boldsymbol{\pi}_{i}^{'}$ to $\boldsymbol{\pi}_{i}^{BBE}$. 

Suppose worker $j$ was matched to task $w$ under $\boldsymbol{\pi}_i^{'}, \boldsymbol{\pi}_{-i}^{'}, \boldsymbol{\pi}_{\mathcal{X}^{c}}$.  When worker $i$ switched to the MTBB trategy from $\boldsymbol{\pi}_{i}^{'}$ there are two possibilities
\begin{itemize}
	\item Task $w$ was not above the task $z$ under  $\boldsymbol{\pi}_{i}^{BBE}$	in which case the  change in strategy does not impact worker $j$
	
	\item Task $w$ was above task $z$ in $\boldsymbol{\pi}_{i}^{BBE}$. Worker $i$ is rejected in favor of another worker. Since the strategies of all the other workers remain fixed the worker $j$ continues to be ranked above all the workers that propose to task $z$. Hence, worker $j$ is matched to task $w$. 
	
\end{itemize}

$$ \hat{U}_{j}(\boldsymbol{\pi}_i^{'}, \boldsymbol{\pi}_{-i}^{'}, \boldsymbol{\pi}_{\mathcal{X}^{c}} ) = \hat{U}_{j}(\boldsymbol{\pi}_i^{BBE}, \boldsymbol{\pi}_{-i}^{'}, \boldsymbol{\pi}_{\mathcal{X}^{c}} )$$

This holds true for all $j\not =i $. Hence, all the workers $j \not=i$ continue to be matched to the same task and derive the same utility when worker $i$ switches. We repeat the same argument one by one for all the other workers and get that 

$$ \hat{U}_{i}(\boldsymbol{\pi}_i^{'}, \boldsymbol{\pi}_{-i}^{'}, \boldsymbol{\pi}_{\mathcal{X}^{c}} ) = U_{i}(\boldsymbol{\boldsymbol{\pi}}^{BBE} | \boldsymbol{m}^F)$$

The above condition is a contradiction. Hence, the workers cannot have strict gains by deviating.

Suppose that the subset consists of only clients. Clients alone cannot gain from deviating as there are no workers to match with.

Suppose that the subset consists of at least one worker and one client. In this case, there will be at least one pair of a client and worker that are matched that gain. If this is the case, then that violates pairwise-stability. 
Hence, in all the three cases we arrive at a contradiction. This shows that it is not possible to have a profitable deviation by a coalition.

\subsection{Proof of Theorem \ref{theorem3}} Before we provide the proof,  it is important to be reminded how we define the uniqueness of the equilibrium. Each equilibrium strategy has a corresponding equilibrium payoff. If for the repeated game that we analyze all the possible equilibrium strategies lead to the same payoff, then we call the equilibrium payoff to be unique. 
In this Theorem, we will assume that the Assumption 1 and 2 hold. We can combine the Assumption 1 and 2 and interpret them together as follows.

From Assumption 1 and 2, we can see that the preference list for all the workers in the MTBB strategy is the same and corresponds to the ranking of the tasks in order of their qualities.  Also, if the workers follow the MTBB strategy, then the ranking of the workers as computed by the planner for every client is the same as well. Specifically, the ranking of the workers is based on the outputs in the assessment phase, where the set of outputs in assessment phase is given as $\{F(k)e_{k}^{max}\}_{k=1}^{N}$. Hence, from now on in this proof when we refer to the ranking of the tasks it is the same as the ranking done by every worker in the MTBB strategy unless stated specifically otherwise. Similarly, when we refer to ranking of workers it is the same as the ranking of the workers based on their maximum outputs computed by planner for every client.

First, we show that there does not exist another equilibrium in which at least one worker $i$ can achieve a higher utility than the utility achieved in the  bang-bang equilibrium (the joint MTBB strategy).  If all the workers play the MTBB strategy, then the matching that is computed at the start of the operational phase after the execution of G-S algorithm is denoted as $\boldsymbol{m}^{BBE}$, where $\boldsymbol{m}^{BBE}[i]$ is the index of the task assigned to worker $i$. Suppose that there exists another equilibrium in which worker $i$ can strictly gain. If  worker $i$ strictly gains in this equilibrium in  comparison to the utility achieved in the bang-bang equilibrium, then it has to be matched to a task that is ranked higher than $\boldsymbol{m}^{BBE}[i]$. Let the task that worker $i$ is assigned to in the new equilibrium be denoted as $\boldsymbol{m}^{BBE}[j]$. In this new equilibrium, we claim that at least one of the workers  that were matched to a task ranked greater than or equal to $\boldsymbol{m}^{BBE}[j]$ in the bang-bang equilibrium will now be matched to a task that is ranked strictly less than its match in the bang-bang equilibrium. Next, we justify this claim.

Consider the set of the workers who were matched to tasks ranked greater than or equal to $\boldsymbol{m}^{BBE}[j]$ in the matching achieved in the operational phase in the  bang-bang equilibrium. Let us denote this set by $\mathcal{U}$. Suppose that the number of workers in this set are $N_1$. In the new equilibrium in which $i$ strictly gains, suppose that every worker in this set is matched to a task that is ranked strictly higher than or equal to its match in the bang-bang equilibrium. First, note that if this supposition is not true, then the claim that there is atleast one worker matched to a task ranked less than its match in the bang-bang equilibrium is already true. Next, we assume that the supposition is true and proceed. Since the worker $i$ is matched to $\boldsymbol{m}^{BBE}[j]$, the workers in $\mathcal{U}$ have to be matched to tasks that are ranked strictly higher than $\boldsymbol{m}^{BBE}[j]$. The total number of tasks that are ranked strictly higher than $\boldsymbol{m}^{BBE}[j]$ are $N_1-1$.
Therefore, if the supposition were true, then  $N_1$ workers have to be matched to at most $N_1-1$ tasks. Hence, it is not possible to assign each of these workers to a strictly higher task (From the Pigeonhole principle). 

Consider the worker that has the highest ranking among all the workers that are assigned to a task that is ranked lower than their corresponding match in the bang-bang equilibrium. Let this worker be worker $k$ and let the task assigned to $k$ in the new equilibrium be $\boldsymbol{m}^{BBE}[l]$. In this new equilibrium, let the worker who is matched to $\boldsymbol{m}^{BBE}[k]$ be worker $r$. Note that the rank of  worker $r$ has to be lesser than the rank of worker $k$. Next, we argue that in this new equilibrium, worker $k$ must have used a strategy different than MTBB. More specifically,  worker $k$ either does not exert maximum effort on at least one task ranked ahead of $\boldsymbol{m}^{BBE}[l]$ in the assessment phase or ranks $\boldsymbol{m}^{BBE}[l]$ ahead of at least one task that was ranked higher in the preference list used in the MTBB strategy. Suppose that this is not the case, which means that  worker $k$ exerts maximum effort on all the tasks ahead of  $\boldsymbol{m}^{BBE}[l]$ and  worker $k$ also ranks all the tasks that were ahead of $\boldsymbol{m}^{BBE}[l]$ to be higher than $\boldsymbol{m}^{BBE}[l]$.


Now since  worker $k$ exerts maximum effort on all the tasks ahead of $\boldsymbol{m}^{BBE}[l]$, it will be ranked ahead of $r$ by $\boldsymbol{m}^{BBE}[k]$  because it has a higher maximum output (rank of $k$ is higher than $k$ in the bang-bang equilibrium). We also know that worker $k$ ranks $\boldsymbol{m}^{BBE}[k]$  ahead of $\boldsymbol{m}^{BBE}[l]$. Therefore, the matching achieved is not stable w.r.t the preferences of the workers and the clients. This is a contradiction as the matching achieved must be stable as we use the G-S algorithm. 
Hence, in the new equilibrium, it must be that the worker must have  either not exerted maximum effort on at least one task ranked ahead of $\boldsymbol{m}^{BBE}[l]$ in the assessment phase or the preference list that it submits must rank $\boldsymbol{m}^{BBE}[l]$ ahead of at least one task that was ranked higher in the preference list in the bang-bang equilibrium. Next, we analyze what happens if  worker $k$ instead uses the MTBB strategy in this case.

In this case,   worker $k$ will approach all the tasks ranked higher than 
$\boldsymbol{m}^{BBE}[l]$ before $\boldsymbol{m}^{BBE}[l]$. We  claim that the worker will be accepted by at least one task ranked higher than or equal to $\boldsymbol{m}^{BBE}[k]$. Suppose that this is not the case, which means no task higher or equal to $\boldsymbol{m}^{BBE}[k]$ accepts $k$. Observe that in the matching achieved in the bang-bang equilibrium, the number of tasks that are ranked higher or equal to $\boldsymbol{m}^{BBE}[k]$ is the same as the number of workers with output greater than or equal to $F(k)e_{k}^{max}$. Based on this observation and the supposition above, it has to be true that at least one of the tasks ranked higher or equal to $\boldsymbol{m}^{BBE}[k]$ accepts a worker with productivity  lower than $F(k)e_{k}^{max}$. Let this task be denoted as $\boldsymbol{m}^{BBE}[q]$. We also know that $\boldsymbol{m}^{BBE}[q]$ is also preferred more by worker $k$ than its current match. Therefore, the matching that is achieved is not stable. This is a contradiction because the matching achieved by the G-S algorithm has to be stable. Hence, it must be true that if  worker $k$ uses MTBB strategy, then it is accepted by a task that is ranked at least as high as $\boldsymbol{m}^{BBE}[k]$.

We assume that not two tasks have the same quality (follows from the assumption that $g$ is strictly increasing). Therefore, $\alpha F(k)^2 g(\boldsymbol{m}^{BBE}[k])-C(k) > \alpha F(k)^2 g(\boldsymbol{m}^{BBE}[l])-C(k)$.
If $\alpha F(k)^2 g(\boldsymbol{m}^{BBE}[k])-C(k)>0$, then  worker $k$ will exert maximum effort in the operational phase and thus deviating to the MTBB strategy  will lead to a  profitable deviation. This is a contradiction as the new equilibrium does not satisfy the incentive compatibility for all the workers. Therefore, $\alpha F(k)^2 g(\boldsymbol{m}^{BBE}[k])-C(k)\leq 0$. In this case,  worker $k$ will have no incentive to exert maximum effort. Thus the deviation cannot be strictly profitable. However, since $\alpha F(k)^2 g(\boldsymbol{m}^{BBE}[k])-C(k)\leq 0$ we can claim in the new equilibrium,  worker $i$ will also exert no effort. If this claim is true, then it will imply that  worker $i$ cannot strictly gain in the new equilibrium, which is a contradiction to the original claim that we can find another equilibrium in which worker $i$ strictly gains. We justify this claim next.

In the new equilibrium,  worker $i$ is matched to task $\boldsymbol{m}^{BBE}[j]$. We know that the rank of $\boldsymbol{m}^{BBE}[k]$ is greater than or equal to $\boldsymbol{m}^{BBE}[j]$, which implies the following 
\begin{equation}
g(\boldsymbol{m}^{BBE}[k])> g(\boldsymbol{m}^{BBE}[j]) 
\label{proof-thm-2-comp-task}
\end{equation}

We also know that the rank of worker $k$ is more than the rank of worker $i$ because in the bang-bang equilibrium worker $k$ is matched to a task that is ranked higher than the task assigned to worker $i$. Therefore,  $F(k)e_k^{max}\geq F(i)e_i^{max}$. From Assumption 1, we can conclude that 

\begin{equation}F(k)e_k^{max}\geq F(i)e_i^{max} \implies F(k)\geq F(i) \implies C(k)\leq C(i) \label{proof-thm-2-comp-prod}
\end{equation}

Based on \eqref{proof-thm-2-comp-task} and \eqref{proof-thm-2-comp-prod}, we have $\alpha F(i)^2 g(\boldsymbol{m}^{BBE}[j])-C(i)\leq 0$,
which implies that the utility achieved by  worker $i$ is $0$. This establishes the claim. Hence, there cannot be another equilibrium in which a worker gains strictly in comparison to the bang-bang equilibrium. 

Next, we argue that there cannot be another equilibrium in which at least one worker gets a strictly lower payoff than in the bang-bang equilibrium.  We develop the proof for this on the same lines as the above. Note that for a worker to have strictly lower utility it has to be that the worker is matched to a task that is ranked strictly lesser than the task the worker is matched to in the bang-bang equilibrium. Also, for the worker to have a strictly lower utility, it has to be true that the worker gets a strictly positive utility from its match in the bang-bang equilibrium. In  this new equilibrium, we define the set of workers who are matched to tasks, which are strictly less in ranking in comparison to their match in the bang-bang equilibrium. From this set, we choose the worker with the highest rank. Let us denote this worker by $s$. Since the worker $s$ has a positive utility from its match in the bang-bang equilibrium it has to be true that for $s$, $\alpha F(s)^2g(\boldsymbol{m}^{BBE}[s])-C(s)>0$. We can show that this worker $s$ must have used a strategy different than MTBB. The proof of this is exactly on the same lines as the one for worker $k$ given above. Based on the above proof for worker $k$ it can also be shown that if  worker $s$ instead uses the MTBB strategy, then it will be matched to a task that has at least the same rank as $\boldsymbol{m}^{BBE}[s]$.  Since $\alpha F(s)^2g(\boldsymbol{m}^{BBE}[s])-C(s)>$ this deviation has to be profitable for worker $s$. Thus in this new equilibrium, incentive compatibility constraints are not satisfied that leads to a contradiction.  
Hence, there will be no equilibrium in which at least one worker gets strictly lower payoff than the bang-bang equilibrium.
We can conclude that in every equilibrium each worker will have the same payoff as in the bang-bang equilibrium.


\subsection{Rearrangement Inequality}

We state the rearrangement inequality next (See \cite{hardy1952inequalities} for details). Suppose $[x_1,x_2,...,x_n]$ and $[y_1,y_2,..., y_n]$ are two ordered lists of numbers, where $x_1\leq x_2 \leq...x_{n-1}\leq x_n$ and $y_1\leq y_2,..., y_{n-1} \leq y_n$. Suppose $\sigma: [1,..,n] \rightarrow [1,..,n]$ is a bijective mapping. We call $\sigma$ a permutation map as it permutes the numbers in the list $[1,..n]$. Rearrangement inequality states that for any permutation map $\sigma$
\begin{equation}
x_1y_1 + x_2y_2 + x_3y_3....x_ny_n \geq x_{\sigma(1) }y_1 + x_{\sigma(2)} y_2 + ..x_{\sigma(n)} y_n
\label{eqn:rearrange}
\end{equation}

\subsection{Proof of Theorem \ref{theorem4}}

Consider a fixed payment rule with parameter $\alpha$. We consider matching rules in which each worker is finally matched to some client (in the limit) and thereafter there is no change in the matching. Given the fixed payment rule, it is easy to check that the only incentive compatible choice for  worker $i$'s effort for task $x$ is 
\begin{equation}e_{i}^{max} I(\alpha F(i)^2g(x)-C(i)\geq 0)
\label{eqn-proof-prop3-effort}
\end{equation}

Therefore, if worker $i$ is matched to task $x$, then the long-run revenue generated by worker $i$ is $F(i)e_{i}^{max} I(\alpha F(i)^2g(x)-C(i) \geq 0)g(x)$. Based on this we can write the expression for the maximum total long-run revenue that can be generated as follows.

\begin{equation}
\max_{\tilde{\boldsymbol{m}}} \sum_{i=1}^{N}F(i)e_{i}^{max} I\Big(\alpha F(i)^2g(\tilde{\boldsymbol{m}}[i])-C(i) \geq 0\Big)g(\tilde{\boldsymbol{m}}[i])
\label{eqn-proof-propn3-max-1}
\end{equation}

%

Next, we simplify the above expression \eqref{eqn-proof-propn3-max-1}. We claim that \eqref{eqn-proof-propn3-max-1} is simplified as follows.

\begin{equation}
\begin{split}
&\max_{\tilde{\boldsymbol{m}}} \sum_{i=1}^{N}F(i)e_{i}^{max} I\Big(\alpha F(i)^2g(\tilde{\boldsymbol{m}}[i])-C(i)\geq 0\Big)g(\tilde{\boldsymbol{m}}[i])\\&=\sum_{i=1}^{N}F(m_i)e_{m_i}^{max} I\Big(\alpha F(m_i)^2g(i)-C(m_i) \geq 0 \Big)g(i)
\end{split}
\label{eqn-proof-propn2-max-3}
\end{equation}

In the above expression \eqref{eqn-proof-propn2-max-3}, $\{F(m_i)e_{m_i}^{max}\}_{i=1}^{N}$ corresponds to the set of values $\{F(i)e_{i}^{max}\}_{i=1}^{N}$ ordered in the increasing order. In the matching in RHS above \eqref{eqn-proof-propn2-max-3}, worker $m_i$ is matched to client $i$. We denote this matching as $\hat{\boldsymbol{m}}$. For consistency, we state that $\hat{\boldsymbol{m}}(m_i)=i, \forall i \in\{1,..,N\}$. 

Next, we establish the above claim by deriving the RHS in \eqref{eqn-proof-propn2-max-3}.

First, we will establish a property that is a consequence of Assumption 2 and Assumption 3.  Define $g_u= \frac{c^{max}}{\Big(f^{min}\Big)^2\alpha}$ and $g_l = \frac{c^{min}}{\Big(f^{max}\Big)^2 \alpha}$.

If worker $i$ is matched to a task $g(y)$ of quality greater than or equal to $g(y)> g_u$, then it will exert maximum effort. To prove this we need to show that the value of the indicator function in \eqref{eqn-proof-prop3-effort} is always one when $g(y)>g_u$.
\begin{equation}
\begin{split}
&I\Big(\alpha F(i)^2g(y)-C(i) \geq 0\Big) \geq I\Big(\alpha (f^{min})^2g(y)-c^{max}\geq 0\Big) \\ &\geq I\Big(\alpha(f^{min})^2g_u-c^{max}\geq 0\Big) =1
\end{split}
\label{eqn-proof-propn3-ineq1}
\end{equation}
If worker $i$ is matched to a task $g(x)$ of quality less than or equal to $g(x)< g_l$, then it will exert no effort. To prove this we need to show that the value of the indicator function in \eqref{eqn-proof-prop3-effort} is always zero when $g(x)<g_l$.
\begin{equation}
\begin{split}
&I\Big(\alpha F(i)^2g(x)-C(i) \geq 0\Big) \leq  I\Big(\alpha (f^{max})^2g(x)-c^{min}\geq 0\Big)\\ & \leq I\Big(\alpha (f^{max})^2g_l-c^{min}\geq 0\Big) =0
\end{split}
\label{eqn-proof-propn3-ineq2}
\end{equation}

Let us assume that there is a $\tilde{\boldsymbol{m}}^{*}$ different than $\hat{\boldsymbol{m}}$, which is  optimal and leads to a strictly higher value for the objective (the total long-run revenue). 
We need to consider the following three cases.

Suppose that there exists at least one task that has a quality more than $g_u$. Therefore, we can find a task denoted as $\hat{j}$ that satisfies the following condition: $g(j)> g_u,  \; \forall j \geq \hat{j}$ and $g(j) < g_l,\;\forall j \leq \hat{j}$. 
For a given matching $\boldsymbol{\tilde{m}}^{*}$, we partition the workers into two sets:  workers that are matched  to tasks with quality greater than or equal to $g(\hat{j})$ and the tasks with quality lesser than  $g(\hat{j})$. Let the two sets for the matching $\hat{\boldsymbol{m}}$ be denoted as $S_1$ and $S_2$, where $S_1$ is the set of workers matched with tasks of quality greater than or equal to $g(\hat{j})$ and $S_2$ is the set of workers matched with tasks with quality lesser than $g(\hat{j})$.  Similarly, the two sets corresponding to the matching $\tilde{\boldsymbol{m}}^{*}$ be $R_1$ and $R_2$. 

Suppose that $R_1$ is not equal to $S_1$. Thus we can conclude that $R_1\cap S_2$ and $R_2\cap S_1$ is non-empty.  Consider a worker $i_1$ from the set $R_1\cap S_2$ and another worker $i_2$ from the set $R_2\cap S_1$. From the definition of the matching $\hat{\boldsymbol{m}}$, we can conclude that  $F(i_1)e_{i_1}^{max}<F(i_2)e_{i_2}^{max}$. In the matching $\tilde{\boldsymbol{m}}^{*}$,  worker $i_1$ is matched to task greater than or equal to $g(\hat{j})$ and  worker $i_2$ is matched to task less than $g(\hat{j})$. 
Suppose that we swap the worker $i_1$ and worker $i_2$ in the matching $\tilde{\boldsymbol{m}}^{*}$. The worker $i_2$ will now exert maximum effort and worker $i_1$ will now exert zero effort (This is due to the property that we established above).
Since $F(i_1)e_{i_1}^{max}<F(i_2)e_{i_2}^{max}$ the total long-run revenue will increase, thus contradicting the fact that $\tilde{\boldsymbol{m}}^{*}$ is optimal. Therefore, the supposition that $R_1$ is not equal to $S_1$ cannot be true. So, we know that $R_1=S_1$ and $R_2=S_2$. Next, we provide the expressions for the total long-run revenues under
$\tilde{\boldsymbol{m}}^{*}$ and $\hat{\boldsymbol{m}}$.
\begin{equation}
\sum_{i\in R_1}F(i)e_i^{max}g(\tilde{\boldsymbol{m}}^{*}[i])
\label{eqn-proof-propn3-simp-1}
\end{equation}

\begin{equation}
\sum_{i\in R_1}F(i)e_i^{max}g(\hat{\boldsymbol{m}}[i])
\label{eqn-proof-propn3-simp-2}
\end{equation}

Since $\tilde{\boldsymbol{m}}^{*}$ is strictly better than $\hat{\boldsymbol{m}}$, it has to be true that the matching $\tilde{\boldsymbol{m}}^{*}$ of the workers within the set $R_1$ is different from $\hat{\boldsymbol{m}}$.
Due to the claim that $\tilde{\boldsymbol{m}}^{*}$ is strictly better than $\hat{\boldsymbol{m}}$, the following has to be true
\begin{equation}\sum_{i\in R_1}F(i)e_i^{max}g(\tilde{\boldsymbol{m}}^{*}[i])>\sum_{i\in R_1}F(i)e_i^{max}g(\hat{\boldsymbol{m}}[i])=\sum_{j=\hat{j}}^{N}F(m_j)e_{m_j}^{max}g(j)\label{Proof-prop3}
\end{equation}

Recall that $\{F(m_i)e_{m_i}^{max}\}_{i=1}^{N}$ corresponds to the set of values $\{F(i)e_{i}^{max}\}_{i=1}^{N}$ ordered in the increasing order. We also know that $\{g(i)\}_{i=1}^{N}$ are task qualities sorted in the increasing order. From rearrangement inequality \eqref{eqn:rearrange}, we know that 
\begin{equation}\sum_{j=\hat{j}}^{N}F(m_j)e_{m_j}^{max}g(j)\geq \sum_{i\in R_1}F(i)e_i^{max}g(\tilde{\boldsymbol{m}}^{*}[i])
\label{eqn-proof-prop3-rearrange}
\end{equation}
The condition above \eqref{eqn-proof-prop3-rearrange} contradicts \eqref{Proof-prop3}.

From the above we get that the set of workers in $R_1$ have to be matched to the same clients by both the matchings $\tilde{\boldsymbol{m}}^{*}$ and $\hat{\boldsymbol{m}}$, which means $\tilde{\boldsymbol{m}}^{*}$ cannot be strictly better than $\hat{\boldsymbol{m}}$.

Observe that the output of our matching rule  is the same as $\hat{\boldsymbol{m}}$ because all the workers rank the clients in the order of their qualities and the all the clients rank the workers based on their maximum outputs. This shows that the total long-run revenue achieved by the proposed mechanism is the same as in \eqref{eqn-proof-propn2-max-3}.

\subsection{Upper Bound on the Performance}	
We write the maximum outputs of workers  sorted in the increasing order as follows $\{F(m_{1})e_{m_{1}}^{max},...,$
$ F(m_{N})e_{m_{N}}^{max}\}$, where $m_x$ is the index of the worker with the $x^{th}$ highest output.
\begin{proposition} 
	\label{prop:expn} If Assumption 2 holds, then the maximum total long-run revenue generated   is $\sum_{x=1}^{N}F(m_x)g(x)e_{m_x}^{max}$.
	
\end{proposition}

\textbf{Proof.} We write the set of  outputs as follows $\{F(1)e_{1}^{max},...,F(N)e_{N}^{max}\}$ and we write the outputs sorted in the increasing order as follows $\{F(m_{1})e_{m_{1}}^{max},...,F(m_{N})e_{m_{N}}^{max}\}$.
Let us first establish the upper bound on  the output.	First, we will compute an upper bound on the total revenue that can be generated  in one period. Clearly, the revenue generated is monotonic in the effort exerted by any worker. Since we are computing the upper bound here we will assume that each worker exerts maximum effort. 	
Each worker $i$ should exert maximum effort $e_{i}^{max}$ otherwise the effort can always be increased to improve the output.  Consider a general matching $\boldsymbol{m}':\mathcal{N} \rightarrow \mathcal{S}$, where $\boldsymbol{m}'[i]$ is the task allocated to worker $i$. 
%
%


We can write the total revenue for this matching $\boldsymbol{m}^{'}$ as follows 	$\sum_{i=1}^{N}F(i)e_{i}^{max}g(\boldsymbol{m}^{'}[i])$. The  inequality given below is a consequence of the rearrangement inequality.
\begin{equation}\sum_{i=1}^{N}F(i)e_{i}^{max}g(\boldsymbol{m}^{'}[i]) \leq \sum_{i=1}^{N}F(m_{i})e_{i}^{max}g(i)\;,\forall \boldsymbol{m}^{'}
\label{eqn-proof-prop2-ubd}
\end{equation}

Therefore, we can also write the following for every matching rule $\boldsymbol{m}$ and  joint strategy $\boldsymbol{\pi}$ as defined in Section \ref{sec2}.
$$\sum_{i=1}^{N}r_{i}(\boldsymbol{h}_{0}^{t}, \boldsymbol{h}_{i}^{t}, \boldsymbol{\pi}_{i}|\boldsymbol{m}) \leq \sum_{i=1}^{N}F(m_{i})e_{m_{i}}^{max} g(i)$$
The above holds true because $r_{i}(\boldsymbol{h}_{0}^{t}, \boldsymbol{h}_{i}^{t}, \boldsymbol{\pi}_{i}|\boldsymbol{m}) = F(i)g(\boldsymbol{m}(\boldsymbol{h}_{0}^{t}))\boldsymbol{\pi}_{i}(\boldsymbol{h}_{i}^{t}) \leq F(i)g(\boldsymbol{m}(\boldsymbol{h}_{0}^{t})[i])e_{i}^{max}$ and $\boldsymbol{m}^{'}=\boldsymbol{m}(\boldsymbol{h}_{0}^{t})$. Note that the upper bound is same for each time slot, the same upper bound continues to hold for the long-run average too. This upper bound is achieved when all the workers cooperate  to maximize the total revenue. Since the revenue is always more than the profit we use the same upper bound for profit (note that it won't be a tight bound).

\subsection{Details of the simulation setup in Section \RomanNumeralCaps{3}.}

In the numerical simulation setup, we will consider the settings where the Assumption 1-2 to hold but we relax Assumption 3. Half of the workers' productivities are independently drawn from a uniform distribution $U\sim[0,w_1]$ and the rest of the workers are drawn independently from a uniform distribution $U\sim [0,w_2]$.  If worker $i$'s productivity is given as $F(i)$, then the cost for exerting effort is defined as $C(i)=C_1-C_2F(i)$. The task qualities are drawn independently from a distribution $t_1+t_2\times U[0,1]$. The maximum effort for all the workers is the same and given as $e^{max}$. 
The linear payment rule is defined as: each client pays the worker a fraction of the revenue generated $\beta \in [0,1]$ to the worker. The quadratic payment rule is defined as: each client $x$ pays the worker $\alpha w^2 g(x)$ for generating $w$ output, where $\alpha\leq \frac{1}{2 \max\{w_1,w_2\} e^{max}}$.
The number of workers are allowed to vary from $10$ to $100$. The other parameters are set as follows $w_1=20$, $w_2=14$, $C_1=2$, $C_2=0.05$, $t_1=2$, $t_2=10$ and the number of draws are set to $10000$.

\subsection{Extensions}

\textbf{General payment and cost functions} Suppose the cost of exerting effort level $e$ for a worker $i$ on a task $x$ is $\boldsymbol{C}(i,x) c(e)$, where $c$ is a convex increasing function of $e$. Suppose that the payment for producing output $w$ on task $x$ is $b(w)g(x)$, where $b$ is an increasing function of $w$. For this cost and payment functions, we continue to use the proposed matching mechanism and we can show that most of the results that we presented extend to this case.  We describe the weakly dominant equilibrium strategy of the worker.  In the assessment phase, the worker exerts maximum effort on every task that it is assigned to. In the reporting phase, the worker ranks the workers based on the utility it expects from the tasks and reports those rankings. In the operational phase, the worker decides the optimal effort level to exert in order to maximize the utility.  The main difference between the equilibrium strategy derived in the main manuscript and here is that the  effort exerted by the workers does not exhibit a bang-bang structure. We can also show that the equilibrium strategy is coalitionally stable (the proof follows the same steps as in Theorem \ref{theorem2}).

\textbf{Client selected payment rules} In this section we expand on the discussion in Section \ref{extsec}. We make the following assumptions.  Clients (task qualities) and workers (productivity, efforts and costs) are drawn i.i.d. from some distribution (known to everyone). For ease of exposition, we assume that the clients use a linear payment rule and the costs for effort are linear in the effort as well. We require the Assumption 1 and 2 to hold as well.  The cost for exerting effort $e_i$ for worker $i$ is $C(i)e_i$.
A client with task of quality $g(x)$ uses a payment rule $\alpha(x) W g(x)$, where $\alpha(x)$ is the fraction that is set by the client, $W$ is the output. As we described in the Section \ref{extsec}, the clients make the same payment per unit output to the workers, which implies $\alpha(x) g(x)$ is the same value for all the clients. This means client with higher task quality pays a lower fraction. Therefore, the client with lowest quality will pay the highest fraction. Suppose the client with lowest quality say client $x$ sets $\alpha(x)=\zeta$, where $\zeta<1$. Based on this the payment rules of the other clients are determined. For instance, a client $y$ will pay $\alpha(y) = \frac{g(x)\zeta}{g(y)}$. Note that $\alpha(y)<1$ since $g(x)$ is lowest quality task.
The expected profit for client $x$ is written as $E[(1-\zeta)F(m_x))g(x) e_{m_x}^{max} I(F(m_x)g(x)\zeta- C(m_x) \geq 0)]$
where $m_x$ is the index of worker that the client is matched with and the expectation is computed using the joint distribution of $C(m_x)$, $F(m_x)$, $e_{m_x}^{max}$. Client $x$ optimizes the above and obtains a $\zeta^{*}$ as the optimal fraction. This determines the payment rules for all the other clients as well, as described above. The client $y$ will pay $\alpha(y) = \frac{g(y)\zeta^{*}}{g(x)}$. 

%

\textbf{Extension to learning by doing.} In the main manuscript, we assumed that the productivities of the workers on the different tasks is fixed. In many cases, such as long-term employment, the workers learn the task and their productivity improves over time \cite{ortega2001job}. Our analysis also extends to this scenario. Instead of assigning each worker to a task for one time slot, we match each worker to a task over multiple time slots, which we refer to as the learning time period. Each worker is assessed based on the output she produces at the end of the  learning time period. The rest of the mechanism is the same. At the end of the assessment phase, the workers submit preferences and workers are ranked based on the output they produce after the  learning time period. G-S algorithm is followed to match the workers and tasks. The equilibrium strategy for the worker remains the same MTBB strategy.

\section{Acknowledgement}
	We would like to acknowledge Professor Gregory Pottie (Department of Electrical and Computer Engineering, UCLA) for valuable comments that helped us improve the paper. We would like to acknowledge the  Office of Naval Research (ONR) and the National Science Foundation (NSF) Award 1524417 for supporting this work. Kartik Ahuja would like to acknowledge the support from the Guru Krupa Fellowship Foundation.

  \bibliographystyle{IEEEtran}
\bibliography{Dynamic_matching_bibfile}
\end{document}